\documentclass[10pt,a4paper]{article}

\usepackage[english]{babel}
\usepackage{graphicx}
\usepackage{hyperref} 
\usepackage{geometry}
\usepackage{amsmath}
\usepackage{amssymb}
\usepackage{amsfonts}
\usepackage{multicol}
\usepackage{slashed}
\usepackage{float}
\usepackage{caption}
\usepackage{subcaption}
\usepackage{cite}
\usepackage{amsthm}
\usepackage{xcolor}
\usepackage{musicography}
\numberwithin{equation}{section}

\newcommand{\M}{\mathcal{M}}
\newcommand{\R}{\mathbb{R}}
\newcommand{\E}{\mathbb{E}}

\newcommand{\p}{\partial}

\def\Xint#1{\mathchoice
	{\XXint\displaystyle\textstyle{#1}}%
	{\XXint\textstyle\scriptstyle{#1}}%
	{\XXint\scriptstyle\scriptscriptstyle{#1}}%
	{\XXint\scriptscriptstyle\scriptscriptstyle{#1}}%
	\!\int}
\def\XXint#1#2#3{{\setbox0=\hbox{$#1{#2#3}{\int}$ }
		\vcenter{\hbox{$#2#3$ }}\kern-.6\wd0}}

\def\dashint{\Xint-}

\def\lowint{\mkern3mu\underline{\vphantom{\intop}\mkern10mu}\mkern-10mu\int}

\title{\bf Analytic Continuation of Stochastic Mechanics}
\author{Folkert~Kuipers$^1$\thanks{E-mail: F.Kuipers@sussex.ac.uk}\\
	$^1${\em Department of Physics and Astronomy, University of Sussex,}\\{\em Brighton, BN1 9QH, United Kingdom}
}

\begin{document}

\maketitle

\begin{abstract}
We study a (relativistic) Wiener process on a complexified (pseudo-)Riemannian manifold. Using Nelson's stochastic quantization procedure, we derive three equivalent descriptions for this problem. If the process has a purely real quadratic variation, we obtain the one-sided Wiener process that is encountered in the theory of Brownian motion. In this case, the result coincides with the Feyman-Kac formula. On the other hand, for a purely imaginary quadratic variation, we obtain the two-sided Wiener process that is encountered in stochastic mechanics, which provides a stochastic description of a quantum particle on a curved spacetime.\\

\end{abstract}

\thispagestyle{empty}
\clearpage
\setcounter{page}{1}

\section{Introduction}
Brownian motion has been at the forefront of physics research ever since the phenomenon, observed by many scientists since the 17th century, was described by Einstein, Schmoluchowski, Langevin, Ornstein, Uhlenbeck, and others, cf. e.g. Ref.\cite{Nelson:1967} for a more detailed historical account. Moreover, it has attracted much attention in the mathematics literature, since the early works on the topic by Wiener, Kolmogorov and L\'evy, and it plays a major role in the stochastic calculus developed by It\^o and Stratonovich.
\par

Nowadays the literature on Brownian motion is rich and extends far beyond its original purpose of describing the motion of pollen suspended in water. In particular, since the introduction of the path integral by Feynman, it has become an important tool in quantum physics. This is mainly due to the Feynman-Kac theorem \cite{FKac}, which made use of the relation between the Euclidean path integral and the Wiener integral. This result became one of the cornerstones of the mathematical foundations of Euclidean quantum field theory, and has been used by several authors as a starting point in attempts to develop a mathematically consistent formulation of Lorentzian quantum field theory, cf. e.g. Refs~\cite{Glimm:1987ng,Albeverio} for reviews.
\par

Later, Parisi and Wu exploited the relation between Brownian motion and Euclidean quantum field theory to develop a framework called stochastic quantization \cite{Parisi:1980ys,Damgaard:1987rr}, which became a very useful computational tool in Euclidean quantum field theory. In recent years, this framework has also been used to relate various string theory inspired models \cite{Dijkgraaf:2009gr,Orlando:2009en,Mansi:2009mz,Heller:2021ofw}.
\par 

Before the work of Parisi and Wu, the notion of stochastic quantization was used by Nelson in the theory of stochastic mechanics \cite{Nelson:1966sp}. This theory, originally proposed by F\'enyes \cite{Fenyes}, serves as an interpretation of quantum mechanics in which quantum mechanics is generated by a two-sided Wiener process \cite{Nelson:1967}. However, later studies of stochastic mechanics were also motivated by the fact that it can be used as a computational framework in quantum theories \cite{Guerra:1981ie} or as a mathematical tool in constructive quantum field theory \cite{Nelson}.
\par

Both the Nelsonian and the Parisi-Wu framework quantize a theory by bringing it in contact with a stochastic background field. However in the Parisi-Wu framework this is done using the one-sided Wiener process, where one considers the forward It\^o differential only, while the Nelsonian approach uses the two-sided Wiener process, where one makes use of both the forward and backward It\^o differentials simultaneously. As the Parisi-Wu framework focuses on the forward It\^o differential, it establishes an equivalence between a Euclidean quantum theory and the equilibrium limit of the stochastic theory. The Nelsonian approach, on the other hand, allows to establish an equivalence between quantum theories and stochastic theories beyond this equilibrium limit. However, as the Nelsonian formulation is more cumbersome than the ordinary theory of Brownian motion, its field theoretic formulation is not as far evolved as the Parisi-Wu formalism.
\par

Since the early work by F\'enyes and Nelson, the theory of stochastic mechanics and its associated stochastic quantization procedure have been extended to include spin~\cite{Nelson,Dankel,Fritsche:2009xu}, to describe processes on (pseudo-)Riemannian manifolds~\cite{Dankel,DohrnGuerraI,DohrnGuerraII,Guerra:1982fn,Nelson,Kuipers:2021jlh,Koide:2019}, and to relativistic theories~\cite{Guerra:1973ck,GuerraRuggiero,Dohrn:1985iu,Marra:1989bi,Guerra:1981ie,Morato:1995ty,Garbaczewski:1995fr,Pavon:2001,Kuipers:2021aok}. Furthermore, field theoretic extensions have been made~\cite{Guerra:1973ck,Guerra:1980sa,Guerra:1981ie,Kodama:2014dba,Morato:1995ty,Garbaczewski:1995fr,Koide:2015,Erlich:2018qfc,NelsonPath,Nelson:2014exa}, but this theory is still in its infancy.
In addition, it is worth noticing that many properties, such as the uncertainty relations, that are sometimes considered to be inherently quantum mechanical naturally arise in stochastic theories \cite{Guerra:1981ie,Nelson,Zambrini,Falco:1983,Golin:1985,Pena,Olavo,Petroni_2000,Gazeau:2019amk,Koide:2012ya}.
\par 

In this paper, we focus on stochastic mechanics of a single particle on a manifold. However, we will do this by reformulating the two-sided Wiener process as a complexified one-sided Wiener process. We will thus study a Brownian motion of a (relativistic) spinless test particle on a complexified (pseudo-)Riemannian manifold. We then find that for a real quadratic variation the one-sided Wiener process, encountered in statistical mechanics, is obtained, while for a purely imaginary quadratic variation the two-sided Wiener process, encountered in stochastic mechanics, is obtained.
\par

This paper is organized as follows: in the next section, we review some aspects of stochastic mechanics; in section 3, we review the connections between stochastic mechanics and generalizations of the Feynman-Kac theorem; in section 4, we discuss the extension of stochastic mechanics to a complex space as proposed in this paper; in sections 5 and 6, we introduce the relativistic stochastic process considered in this paper and the manifold on which this process is defined; in section 7, we discuss the variational equations that govern the stochastic process; in sections 8, 9 and 10, we then derive three different formulations for the diffusion problem; finally, in section 11, we conclude. Furthermore, in appendix A, we summarize our results for the non-relativistic case; in appendix B, we review the basics of stochastic integration and appendix C contains calculations of conditional expectations that are necessary to derive our results. Throughout the paper we work in Planck units, i.e. $\hbar=1$, $c=1$, $G=1$ and $k_B=1$. Moreover, we work in the $(-+++)$ signature convention.

\section{Stochastic Mechanics}\label{sec:StochMech}
In order to illustrate the ideas governing stochastic mechanics, we will start with the discussion of a single scalar non-relativistic particle with mass $m$ moving on the configuration space $\R^n$.
\par 

In classical mechanics the motion of such a particle is governed by the Euler-Lagrange equations. These can be derived using a variational principle from the action
\begin{equation}
	S = \int_0^T L(x,v) \, dt.
\end{equation}
with Lagrangian $L$. Given some initial conditions $(x,v)(0)=(x_0,v_0)$ one then obtains a unique solution $(x,v)(t):\mathcal{T}\rightarrow \R^{2n}$ with $\mathcal{T}=[0,T]$.
\par 

We now make the additional assumption that the particle moves through some randomly fluctuating background field or quantum vacuum.
In order to describe the stochastic dynamics induced by this background field, we promote the configuration space to a measurable space $\left(\R^n,\mathcal{B}(\R^n)\right)$ with Borel sigma algebra. Moreover, we introduce the probability space $(\Omega,\Sigma,\mathbb{P})$ and study random variables $X:(\Omega,\Sigma,\mathbb{P})\rightarrow\left(\R^n,\mathcal{B}(\R^n),\mu\right)$ with $\mu=\mathbb{P} \circ X^{-1}$.
\par 

In addition, we introduce filtrations $\{\mathcal{P}_t \}_{t\in\mathcal{T}}$ and $\{\mathcal{F}_t \}_{t\in\mathcal{T}}$, which we call the past and future filtration. The past filtration $\{\mathcal{P}_t \}_{t\in\mathcal{T}}$ is an ordered set that is increasing, i.e. $\emptyset \subseteq \mathcal{P}_s \subseteq \mathcal{P}_t \subseteq \Sigma \quad \forall s<t\in\mathcal{T}$, and right-continuous, i.e. $\mathcal{P}_t = \cap_{\epsilon>0} \mathcal{P}_{t+\epsilon}$. The future filtration, on the other hand, is an ordered set that is decreasing, i.e. $\emptyset \subseteq \mathcal{F}_s \subseteq \mathcal{F}_t \subseteq \Sigma \quad \forall s>t\in\mathcal{T}$, and left-continuous, i.e. $\mathcal{F}_t = \cap_{\epsilon>0} \mathcal{F}_{t-\epsilon}$.
\par

We study stochastic processes, i.e. families of random variables $\{X_t: t\in\mathcal{T}\}$, adapted to such filtrations. More precisely, in stochastic mechanics one studies processes that are semi-martingales with respect to both the past and future filtration, which means that the process can locally be decomposed as
\begin{align}
	X_t &= C^+_t + M^+_t,\nonumber\\
	X_t &= C^-_t + M^-_t,
\end{align}
where $C^+_t$ is a continuous local c\`adl\`ag process with finite variation, $C^-_t$ is a continuous local c\`agl\`ad process with finite variation, and $M^\pm_t$ are continuous local martingales with respect to the past and future filtration respectively. Hence, they satisfy the martingale property
\begin{align}
	\E\big[ M^+_t \big| \mathcal{P}_s \big] &= M^+_s \qquad \forall\, s<t \in \mathcal{T},\nonumber\\
	\E\big[ M^-_t \big| \mathcal{F}_s \big] &= M^-_s \qquad \forall\, s>t \in \mathcal{T}.
\end{align}
\par

One can then introduce two velocities:
\begin{align}
	v_+^i(X_t,t) &= \lim_{h\rightarrow 0} \frac{1}{h} \, \E \left[X^i_{t+h} - X^i_t \Big| \mathcal{P}_t \right],\nonumber\\
	v_-^i(X_t,t) &= \lim_{h\rightarrow 0} \frac{1}{h} \, \E \left[X^i_t - X^i_{t-h} \Big| \mathcal{F}_t \right].
\end{align}
where the forward velocity is a conditional expectation with respect to the past filtration, and the backward velocity is a conditional expectation with respect to the future filtration. In addition, one can construct the objects\footnote{We note that this definition deviates by a factor $2$ compared to the definition used in our previous works \cite{Kuipers:2021jlh,Kuipers:2021aok}.}
\begin{align}
	v_+^{ij}(X_t,t) 
	&= \lim_{h\rightarrow 0} \frac{1}{h} \, \E \left[ \left(X^i_{\tau+h} - X^i_t \right) \left(X^j_{\tau+h} - X^j_t\right) \Big| \mathcal{P}_t \right],
	\nonumber\\
	v_-^{ij}(X_t,t) 
	&= - \lim_{h\rightarrow 0} \frac{1}{h} \, \E \left[ \left(X^i_t - X^i_{t-h}\right) \left(X^j_t - X^j_{t-h}\right) \Big| \mathcal{F}_t \right].
\end{align}
We point out that these objects vanish for any deterministic motion, but are non-vanishing for stochastic processes. Finally, it is customary to define the objects
\begin{align}
	v &= \frac{1}{2} \left( v_+ + v_-\right),\nonumber\\
	u &= \frac{1}{2} \left( v_+ - v_-\right),
\end{align}
where $v$ is called the current velocity and can be associated with the particle itself. $u$ is the osmotic velocity and is associated with the background field.
\par

The stochastic quantization procedure as applied in stochastic mechanics and reviewed in Ref.~\cite{Nelson} then states that the motion of a quantum particle minimizes a stochastic action
\begin{equation}
	S = \E\left[ \int L(X,V_+,V_-) \, dt\right]
\end{equation}
using a stochastic variational principle \cite{Yasue}. The resulting equations are the stochastic Euler-Lagrange equations, which is a set of stochastic differential equations. On top of this, one must fix the stochastic fluctuations of the background field by imposing the background hypothesis 
\begin{equation}
	v_+^{ij} = \frac{\hbar}{m} \, \delta^{ij} = - v_-^{ij}.
\end{equation}
\par 

Given these conditions, one can show that, if one defines a wave function
\begin{equation}\label{eq:WaveFunctionNelson}
	\Psi = e^{\frac{1}{\hbar}\left(R + {\rm i} \, S \right)}
\end{equation}
with
\begin{equation}
	R = \frac{\hbar}{2} \ln\rho
\end{equation}
and $\rho(x,t)$ the probability density of finding the particle at a position $x$ at time $t$, then the stochastic Hamilton-Jacobi equations impose that $\Psi$ satisfies the Schr\"odinger equation. Moreover, the Born rule
\begin{equation}
	\rho = |\Psi|^2
\end{equation}
is automatically satisfied.
\par

We conclude this section by pointing out several issues encountered in stochastic mechanics. The first is that the wave function \eqref{eq:WaveFunctionNelson} is not well-defined if the configuration space is not simply connected \cite{Wallstrom:1988zf,WallstromII}. However, this issue can be resolved by considering the universal cover of the configuration space \cite{Nelson}.
\par

The second issue is that there is an ambiguity in the construction of the stochastic Lagrangian $L(X,V_+,V_-)$ from a classical Lagrangian $L_c(x,v)$. One prescription was given by  Guerra and Morato \cite{Guerra:1982fn} and states that
\begin{equation*}
	L(X,V_+,V_-) = L_c(X,V_+)
\end{equation*}
Another prescription was given by Yasue \cite{Yasue:1981wu} and uses
\begin{equation*}
	L(X,V_+,V_-) = \frac{1}{2} \left[ L_c(X,V_+) + L_c(X,V_-)\right]
\end{equation*}
It was then shown by Zambrini in Ref.~\cite{Zambrini} that Yasue's prescription should be favored, as it respects gauge invariance.
A third prescription was given by Pavon \cite{Pavon:1995April,Pavon:1995June,Pavon:1996} and makes use of a complex velocity field
\begin{equation}\label{eq:ComplexVelPavon}
	v_q = v - {\rm i} \, u
\end{equation}
such that one can define
\begin{equation*}
	L(X,V_+,V_-) = L_c(X,V_q).
\end{equation*}
\par

Finally, it is worth emphasizing that stochastic mechanics is an attempt to construct a well-defined classical probabilistic formulation of quantum mechanics. In doing so, it introduces the stochastic fluctuations of the covariant background field or quantum vacuum as a fundamental law of nature. It does not posit the existence of any deterministic hidden variables, and is therefore not in contradiction with the Bell experiments. This is in contrast with the theory of Brownian motion of a pollen suspended in water, where the stochastic theory is known to be an effective theory replacing a more fundamental theory that takes into account the motion of the water molecules hitting the pollen.
\par

As an interpretation of quantum mechanics, stochastic mechanics is thus agnostic about the question whether God is playing dice in the probability space or whether there exists a more fundamental theory from which the stochastic theory can be derived. Nevertheless, in the literature, there are proposals to provide such a more fundamental theory. An example is provided by Calogero's conjecture \cite{Calogero:1997,Calogero:2004} in which the quantum vacuum is caused by the chaotic behavior induced by the gravitational interaction between all matter in the universe.

\section{Stochastic Mechanics and the Feynman-Kac Theorem}
The Feynman-Kac theorem states\footnote{We present an elementary form of the Theorem. Extensions beyond the formula presented here are known.} that, given the real diffusion equation
\begin{equation}\label{eq:RealDiffusion}
	\frac{\p}{\p t} \Psi(x,t) 
	= 
	- \left[ 
	\frac{\alpha}{2} \, \delta^{ij} \, \p_i \p_j 
	+ v^i(x,t) \, \p_i
	- \mathfrak{U}(x,t)
	\right] \Psi(x,t)
\end{equation}
with $\alpha>0$, $x\in\R^n$ and $t\in[0,T]$ subjected to the terminal condition
\begin{equation}
	\Psi(x,T) = u(x),
\end{equation}
the solution can be written as the conditional expectation
\begin{equation}
	\Psi(x,t) = \E\left[\exp\left(-\int_t^T \mathfrak{U}(X_s,s) \, ds \right) u(X_T) \Big| X_t = x \right]
\end{equation}
for the It\^o process defined by the stochastic differential equation
\begin{align}\label{eq:ItoProc}
	dX^i_t &= v^i(X_t) \, dt + dM^i_t,\nonumber\\
	d[X^i,X^j]_t &= \alpha \, \delta^{ij} \, dt,
\end{align}
with $M_t$ a local martingale with respect to an increasing filtration. Furthermore, $[X^i,X^j]$ denotes the quadratic covariation defined by
\begin{align}
	[X^i,X^j]_t &= \int_0^t d[X^i,X^j]_t \nonumber\\
	&= \lim_{k\rightarrow\infty} \sum_{[t_l,t_{l+1}]\in\pi_k} 
	\big(X^i_{t_{l+1}} - X^i_{t_l} \big) \big(X^j_{t_{l+1}} - X^j_{t_l} \big)\, ,
\end{align}
where $\pi_k$ is a partition of $[0,t]$ into $k$ intervals. We note that the square bracket is the standard notation for both the quadratic covariation and the commutator. In this paper the square bracket always denotes a quadratic covariation, as we do not encounter commutators. Moreover, we point out that the quadratic covariation vanishes for any deterministic process, but is non-vanishing for stochastic processes.
\par

It was suggested by by Gelfand and Yaglom \cite{Gelfand}, that a similar relation could exist for the Schr\"odinger equation
\begin{equation}\label{eq:ComplexDiffusion}
	{\rm i} \, \frac{\p}{\p t} \Psi(x,t) 
	= 
	- \left[ 
	\frac{\alpha}{2} \, \delta^{ij} \, \p_i \p_j 
	+ v^i(x,t) \, \p_i
	- \mathfrak{U}(x,t)
	\right] \Psi(x,t).
\end{equation}
However, soon after, it was pointed out by Cameron and Daletskii \cite{Cameron,Daletskii,Albeverio} that a straightforward generalization does not exist, as the complex measure necessary to construct such an equivalence will have an infinite total variation.
\par

Later, Pavon \cite{Pavon:2000} showed that such a relation could still be established, if one considers, instead of the process \eqref{eq:ItoProc}, a two-sided Wiener process defined by
\begin{align}\label{eq:ItoProcF}
	d_+X^i_t &= v_+^i(X_t) \, dt + d_+M^i_t,\nonumber\\
	d_+[X^i,X^j]_t &= \alpha \, \delta^{ij} \, dt
\end{align}
and
\begin{align}\label{eq:ItoProcB}
	d_-X^i_t &= v_-^i(X_t) \, dt + d_-M^i_t,\nonumber\\
	d_-[X^i,X^j]_t &= \alpha \, \delta^{ij} \, dt
\end{align}
with $d_+$ a forward It\^o differential and $d_-$ a backward It\^o differential. The important difference between the description given in eq.~\eqref{eq:ItoProc} and the one described by eqs.~\eqref{eq:ItoProcF} and \eqref{eq:ItoProcB} is that the latter focuses on both the forward and backward It\^o differential simultaneously, while the former uses the forward differential only, cf. Ref.~\cite{Nelson:1967} for more detail.
\par 

In deriving this result Pavon builds on earlier work \cite{Pavon:1995April,Pavon:1995June,Pavon:1996} using the complex velocity given in eq.~\eqref{eq:ComplexVelPavon}. Therefore, in contrast to the earlier works \cite{Gelfand,Cameron,Daletskii}, Pavon did not only complexify the measure, but also the underlying degrees of freedom: the velocity of the process. In this work we go one step further and also complexify the position of the process. This will be discussed in more detail in the next section. 

\section{Complexification of the configuration space}\label{sec:Complexification}

Although the two-sided Wiener process \cite{Nelson:1967} has been studied extensively in stochastic mechanics, the study of the one-sided Wiener process, where one focuses on only one time generator, is more common in the stochastic literature. In this section, we show how stochastic mechanics can be rewritten in terms of the one-sided process by complexifying the configuration space.
\par

Before doing so, we point out that complex extensions of stochastic mechanics have been considered earlier in the literature by Rosenbrock, cf. Refs.~\cite{Rosenbrock:1986,Rosenbrock:1995,Rosenbrock:1997,Rosenbrock:1999,Rosenbrock:2000}. In contrast to these works, which develop the theory within the framework of stochastic optimal control theory, we develop the stochastic Euler-Lagrange equations on (pseudo-)Riemannian manifolds.
\par

We provide another motivation for our approach by recalling the L\'evy characterization of a Brownian motion \cite{Levy}, which states that a continuous $\R^n$-valued martingale process $M_t$ is a $n$-dimensional Brownian motion if and only if the quadratic variation is given by
\begin{equation}\label{LevyMartingale}
	d[M^i,M^j]_t =  \delta^{ij} \, dt.
\end{equation}
\par

This can easily be generalized to a $n$-dimensional scaled Brownian motion with drift. Indeed, this is a semi-martingale process $X_t= C_t + \sqrt{\alpha}\, M_t$ such that $C_t$ is a c\`adl\`ag process with finite variation and $M_t$ is a martingale satisfying \eqref{LevyMartingale} or equivalently
\begin{equation}\label{eq:GaussQuadVar}
	d[X^i,X^j]_t = \alpha \, \delta^{ij} \, dt,
\end{equation}
where $\alpha$ is a positive real constant.
\par

A similar characterization exists for the compensated Poisson process, cf. e.g. Ref.~\cite{Emery}, which is characterized by the structure relation
\begin{equation}
	d[X^i,X^j]_t = \alpha \, \delta^{ij} \, dt + \delta^{ij} \, c_k \, dX_t^k,
\end{equation}
where $c_k$ is a constant covector.
\par

We thus see that the stochastic behavior of various classes of stochastic processes is characterized by a structure relation for the quadratic variation of the process. This is reminiscent to the characterization of quantum mechanics by a structure relation for the commutators of operators in the canonical commutation prescription.
Based on this observation, one can hope that there also exists a quantum structure relation on the quadratic variation. 
\par

By considering the complex diffusion equation
\begin{equation}\label{eq:DiffusionEq}
	\alpha \, \frac{\p}{\p t} \Psi(x,t) = \left( \frac{\alpha^2}{2m} \, \frac{\p}{\p x_i} \frac{\p}{\p x^i}  + \mathfrak{U}(x) \right) \Psi(x,t)
\end{equation}
which reduces to the heat equation with diffusion constant $\frac{\hbar}{2m}$ for $\alpha=\hbar$ and to the Schr\"odinger equation for $\alpha={\rm i \, \hbar}$, and by noting that the Feynman-Kac theorem allows to associate a structure relation 
\begin{equation}
	m \, d[X^i,X^j]_t = \hbar \, \delta^{ij} \, dt
\end{equation}
to the heat equation, one can argue that a quantum structure relation must be of the form
\begin{equation}\label{eq:GaussQuadVarComplex}
	m \, d[X^i,X^j]_t = {\rm i} \, \hbar \, \delta^{ij} \, dt.
\end{equation}
However, since the right hand side is not a positive definite tensor, there does not exist a real semi-martingale $X$ satisfying this relation.
\par

As discussed in previous sections, this issue is circumvented in stochastic mechanics by considering the two time generators $d_+$ and $d_-$ simultaneously. This allows to construct the two-sided Wiener process $X$, which satisfies the condition 
\begin{align}
	m \, d_+[X^i,X^j]_t &= \delta^{ij} \, dt,\nonumber\\
	m \, d_-[X^i,X^j]_t &= \delta^{ij} \, dt.
\end{align}
It is well-established that this process generates quantum mechanics of a spin-0 particle with mass $m$, cf. e.g. \cite{Nelson,Zambrini,Guerra:1981ie} for reviews.
\par

In this paper, we advocate a slightly different route. Instead, of considering two time generators $d_\pm$ simultaneously, we will consider only one. However, we complexify our space $\R^n$ to the complex space $\mathbb{C}^n$ and analytically continue all functions defined on this space including the wave function. 
\par

In order to introduce stochastic dynamics, we will promote the complex configuration space to a measurable space $\left(\mathbb{C}^n,\mathcal{B}(\mathbb{C}^n)\right)$ with Borel sigma algebra. Moreover, we introduce the probability space $(\Omega,\Sigma,\mathbb{P})$ and study random variables $Z:(\Omega,\Sigma,\mathbb{P})\rightarrow\left(\mathbb{C}^n,\mathcal{B}(\mathbb{C}^n),\mu\right)$ with $\mu=\mathbb{P} \circ Z^{-1}$. More precisely, we study stochastic processes, i.e. families of random variables $\{Z_t: t\in\mathcal{T}\}$. 
\par

We assume the stochastic processes $Z_t$ to be complex continuous semi-martingale processes adapted to an increasing filtration $\{\mathcal{P}_t \}_{t\in\mathcal{T}}$. These are processes $Z_t = X_t + {\rm i} \, Y_t$ such that $X_t$ and $Y_t$ are continuous real semi-martingales, i.e, they can be decomposed as
\begin{align}\label{DoobMeyer}
	X^i_t &= C^i_{x,t} + M^i_{x,t},\nonumber\\
	Y^i_t &= C^i_{y,t} + M^i_{y,t},
\end{align}
where $C^i_{x,t}$ and $C^i_{y,t}$ are continuous c\`adl\`ag processes with finite variation and $M^i_{x,t}$ and $M^i_{y,t}$ are continuous martingales satisfying the martingale property
\begin{align}
	\E\big[ M^i_{x,s} \big| \mathcal{P}_t \big] &= M^i_{x,t} \qquad \forall\, t<s \in \mathcal{T},\nonumber\\
	\E\big[ M^i_{y,s} \big| \mathcal{P}_t \big] &= M^i_{y,t} \qquad \forall\, t<s \in \mathcal{T}.
\end{align}
In the remainder of the paper, we will use a shorthand notation for conditional expectation values:
\begin{equation}
	\E_{t} \big[f(Z_s) \big] := \E\big[ f(Z_s) \big| \mathcal{P}_t \big].
\end{equation}
Using this notation, the C\`adl\`ag process $C_t= C_{x,t} + {\rm i} \, C_{y,t}$ can be reconstructed as
\begin{equation}
	C^i_t - C^i_0 
	= 
	\lim_{h\rightarrow 0} \int_{0}^{t} \frac{1}{h}\, \E_{s} \left[Z^i_{s+h} - Z^i_s \right]  ds
\end{equation}
and the angle bracket process is given by
\begin{equation}
	\langle Z^i , Z^j \rangle_\tau
	- \langle Z^i , Z^j \rangle_0
	\nonumber\\ 
	= 
	\lim_{h\rightarrow 0} \int_{0}^{\tau} \frac{1}{h} \, 
	\E_{s} \Big[  
	\Big(Z^i_{s+h} - Z^i_s \Big) 
	\Big(Z^j_{s+h} - Z^j_s \Big) 
	\Big] ds.
\end{equation}
This is the compensator for the quadratic variation, i.e., the process
\begin{equation}
	[Z^i , Z^j]_t - \langle Z^i , Z^j \rangle_t
\end{equation}
is a local martingale with respect to $\{\mathcal{P}_t\}_{t\in\mathcal{T}}$. For more detail on the theory of semi-martingales we refer to e.g. the appendix of Ref.~\cite{Emery} and references therein.
\par 

We can now impose a condition on the quadratic variation of the form
\begin{equation}\label{eq:StrucRel}
	m \, d[Z^i,Z^j]_t = \alpha \, \delta^{ij} \, dt,
\end{equation}
where $\alpha\in\mathbb{C}$. Similarly, for the complex conjugate process $\bar{Z}$, we impose
\begin{align}
	m \, d[\bar{Z}^i,\bar{Z}^j]_t &= \bar{\alpha} \, \delta^{ij} \, dt,\nonumber\\
	m \, d[Z^i,\bar{Z}^j]_t &= (|\alpha| + \beta) \, \delta^{ij} \, dt
\end{align}
with $\beta\in[0,\infty)$. If we use the polar decomposition $\alpha=\rho\, e^{{\rm i} \phi}$, we find that this is equivalent to
\begin{align}
	m \, d[X^i,X^j]_t &= \frac{\beta + \rho \, (1 + \cos\phi)}{2} \,  \delta^{ij} \, dt,\nonumber\\
	m \, d[Y^i,Y^j]_t &= \frac{\beta + \rho \, (1 - \cos\phi)}{2} \,  \delta^{ij} \, dt,\nonumber\\
	m \, d[X^i,Y^j]_t &= \frac{\rho \, \sin\phi}{2} \, \delta^{ij} \, dt.
\end{align}
Hence, the quadratic variation of both $X$ and $Y$ is positive definite, as required for their existence.
We note that we recover the Gaussian increments process defined by eq.~\eqref{eq:GaussQuadVar} for $(\alpha,\beta)=(1,0)$, while we recover a relation similar to the one suggested in eq.~\eqref{eq:GaussQuadVarComplex} for $(\alpha,\beta)=({\rm i},0)$.
\par

In the remainder of the paper, we study semi-martingale processes that satisfy the structure relation \eqref{eq:StrucRel} with $\beta=0$ for a general $\alpha\in\mathbb{C}$ in more detail. More precisely, we study these processes using the stochastic quantization procedure discussed in Ref.~\cite{Nelson}, and show, using the Hamilton-Jacobi formalism, that such processes can be associated to the complex diffusion equation \eqref{eq:DiffusionEq}, where $\Psi(x,t)$ is analytically continued to $\Psi(z,t)$.
We will perform our analysis in the more complicated setting where the particle is relativistic and moves on a curved spacetime. 
\par 

Before moving on, we must give a physical interpretation to the additional dimensions that are introduced by extending the configuration space from $\R^n$ to the complex space $\mathbb{C}^n$. For this, we use that the introduction of a complex velocity $v_q=v-{\rm i}\,u$ in stochastic mechanics, as discussed in Refs.~\cite{Pavon:1995April,Pavon:1995June,Pavon:1996}, naturally follows from the fact that there exist two well-defined conditional derivatives $v_\pm$ for any $\R^n$-valued finite energy diffusion process, and the requirement that the classical velocity is recovered in the semi-classical limit. Moreover, the current velocity $v$ can be associated with the velocity of the particle itself, while the osmotic velocity $u$ can be associated with the velocity of the background field.
\par

We can adopt this interpretation for the complex velocity $W=V+{\rm i}\, U$ and call $V$ the current velocity of the matter and $U$ the osmotic velocity of the background. We can then extend this interpretation to the complex position $Z=X+{\rm i}\, Y$ and associate the position $X$ to the particle itself, while $Y$ is the position of an associated particle in the background field.
\par

The fluctuations of the covariant background field in stochastic mechanics can be regarded as fluctuations of spacetime itself or as a quantum foam \cite{Wheeler:1955zz}. Adopting this point of view, we see that both the spacetime itself and matter defined on the spacetime evolve. Furthemore, the evolution of the matter is encoded in the real coordinate $X$, while the evolution of spacetime is encoded in the complex coordinate $Y$.

\section{The Geometry}
We will generalize the discussion in previous sections to the context of relativistic particles on Lorentzian manifolds. For this, we consider a set $\mathcal{T}=[0,T]$, a real $(n=d+1)$-dimensional Lorentzian manifold $(\M,g)$, and trajectories $x(\tau):\mathcal{T}\rightarrow\M$.
\par 

We intend to superpose stochastic dynamics on these trajectories. However, stochastic dynamic violates the Leibniz rule, as stochastic processes have a non-vanishing quadratic variation. As a consequence, diffeomorphism invariance of stochastic theories defined on this manifold is broken. In this paper, we resolve this issue using the second order geometry framework as developed by Schwartz, Meyer and Emery \cite{Schwartz,Meyer,Emery}.
\par 
 
The most important aspect of the second order geometry framework is that all tangent spaces $T_x\M$ are extended to second order tangent spaces $T_{2,x}\M$. In a local coordinate chart, second order vectors can be expressed as\footnote{We slightly deviate from Refs.~\cite{Emery,Kuipers:2021jlh}, as we have introduced a factor $\frac{1}{2}$ in the second order part of the vector.}
\begin{equation}
	v = v^\mu \, \p_\mu + \frac{1}{2} \, v^{\mu\nu} \, \p_\mu\p_\nu,
\end{equation}
where $v^\mu \p_\mu \in T_x\M\subset T_{2,x}\M$ represents the first order part and $v^{\mu\nu} \p_\mu\p_\nu$ the second order part. This second order part can be mapped bijectively onto a symmetric bilinear first order tensor, which in turn can be mapped bijectively onto the quadratic variation of the process $X_t$.\footnote{cf. Theorem 3.8 and Proposition 6.13 in Ref.~\cite{Emery}}
\par 

When regarded as part of a second order vector, the first order vector $v^\mu \p_\mu \in T_x\M$ no longer transforms in a covariant manner. However, one can construct the objects
\begin{align}
	\hat{v}^\mu &= v^\mu + \frac{1}{2} \, \Gamma^\mu_{\sigma\kappa} v^{\sigma\kappa},\nonumber\\
	\hat{v}^{\nu\rho} &= v^{\nu\rho},
\end{align}
which both transform covariantly. Diffeomorphism invariance of the physical theory can then be restored by replacing all vectors $v^\mu$ with their covariant expression $\hat{v}^\mu$.
\par 

For a more complete exposition of the material, we refer to the works of Schwartz, Meyer and Emery \cite{Schwartz,Meyer,Emery}. We note that the construction of a diffeomorphism invariant theory of stochastic mechanics was already studied extensively, cf. e.g. Refs.~\cite{DohrnGuerraI,DohrnGuerraII,Nelson}. Recently, we have translated and extended parts of these results into the second order geometry language \cite{Kuipers:2021jlh}.
\par

As a final step, we need to complexify the manifold to $\M^\mathbb{C}=\M\otimes \mathbb{C}$. Similarly, the tangent spaces are complexified, such that we obtain a first and second order tangent bundle
\begin{align*}
	(T\M)^\mathbb{C} &= T\M \otimes \mathbb{C} = T^{1,0}\M \oplus T^{0,1}\M,\\
	(T_2\M)^\mathbb{C} &= T_2\M \otimes \mathbb{C} = T_2^{1,0}\M \oplus T_2^{0,1}\M.
\end{align*}

\section{The Stochastic Process}
We will now introduce stochastic dynamics as described in section \ref{sec:Complexification}. We must thus promote the complex manifold to a measurable space $\left(\M^\mathbb{C},\mathcal{B}(\M^\mathbb{C})\right)$ and study continuous semi-martingale processes on this manifold. As we are working on manifolds the Doob-Meyer decomposition given in eq.~\eqref{DoobMeyer} is only valid locally.
Thus for every coordinate chart $\chi:U\rightarrow V$ with $U\subset\M^\mathbb{C}$ and $V\subset\mathbb{C}^n$ the processes $Z^\mu=\chi^\mu(Z)$ are continuous semi-martingales, i.e, they can locally be decomposed uniquely as
\begin{align}
	Z^\mu_\tau &= C^\mu_{\tau} + M^\mu_{\tau}.
\end{align}
with $C^\mu_\tau$ a continuous local c\`adl\`ag process with finite variation and $M^\mu_\tau$ is a continuous local martingale satisfying the martingale property
\begin{align}
	\E\big[ M^\mu_\tau \big| \mathcal{P}_t \big] &= M^\mu_t \qquad \forall\, t<\tau \in \mathcal{T}.
\end{align}
We can associate a velocity to the process, which is given by
\begin{align}
	w^\mu(Z_\tau,\tau) 
	&= 
	\lim_{h\rightarrow 0} \frac{1}{h} \E_{\tau} \left[Z^\mu_{\tau+h} - Z^\mu_\tau \right],
\end{align}
and is the first order part of a second order vector field $w^{^\mu_{\nu\rho}}$.
\par 

The process $Z$ can be lifted to the tangent bundle, yielding a stochastic process $(Z_\tau,W_\tau)$ on the second order holomorphic tangent bundle $T_2^{1,0}\M$, which can be decomposed into the real processes $X_\tau,Y_\tau$ and $V_\tau,U_\tau$ such that
\begin{align}
	Z_\tau &= X_\tau + {\rm i} \, Y_\tau,\nonumber\\
	W_\tau &= V_\tau + {\rm i} \, U_\tau.
\end{align}
Furthermore, for the natural projection $\pi:T_2^{1,0}\M\rightarrow \M$, $\pi(Z_\tau,W_\tau)=Z_\tau$, and
\begin{align}
	W^\mu_\tau \, d\tau &= \circ \, dZ^\mu_\tau,\\
	W^{\nu\rho} \, d\tau &= d[Z^\nu,Z^\rho]_\tau.\label{eq:VelProc2}
\end{align}
An immediate consequence is that, cf. eqs.~\eqref{eq:CondExpDerQvar} and~\eqref{eq:CondExpDer},
\begin{align}
	w^\mu(Z_\tau,\tau) &= \E_\tau\left[W^\mu_\tau \right],\nonumber\\
	w^{\nu\rho}_\tau(Z_\tau,\tau) &= \E_\tau\left[W^{\nu\rho}_\tau \right].
\end{align}
We remark that, as discussed in the previous section, the object $w^{^\mu_{\nu\rho}}$ is not covariant. However, one can obtain a covariant formulation given by
\begin{align}
	\hat{w}^\mu &= w^\mu + \frac{1}{2} \, \Gamma^\mu_{\sigma\kappa} w^{\sigma\kappa},\nonumber\\
	\hat{w}^{\nu\rho} &= w^{\nu\rho}.
\end{align}
\par

As discussed in section \ref{sec:Complexification}, we would like to fix the quadratic variation of the processes by
\begin{equation*}
	d[Z^\mu,Z^\nu]_\tau
	= \alpha \, \lambda\, g^{\mu\nu}(Z_\tau) \, d\tau,
\end{equation*}
where $\alpha\in \mathbb{C}$ and $\lambda$ is a dimensionful constant characterizing the particle. However, as the metric tensor is not positive definite, there exists no semi-martingale $Z$ satisfying this relation. In the literature, there exists several resolutions for this issue. One is given in Ref.~\cite{Pavon:2001}, where the stochastic dynamics is restricted to the spatial coordinates, and the time coordinate is a stopping time associated to this spatial process.
\par 

Here, we will follow the solution proposed in Ref.~\cite{Guerra:1973ck,GuerraRuggiero,Dohrn:1985iu}, as it can easily be embedded in the second order geometry framework such that general covariance is preserved \cite{Kuipers:2021aok}. In this approach one performs a transformation on the metric such that a Euclidean (Brownian) metric $g_{\rm E}$ is obtained from the Lorentzian (kinetic) metric $g$:
\begin{equation}
	g_{\rm E}^{\mu\nu} = g^{\mu\nu} + 2\, u^\mu u^\nu.
\end{equation}
with $u$ such that $g_{\mu\nu}u^\mu u^\nu = -1$. The quadratic variation can then be fixed as
\begin{equation}
	d[Z^\mu,Z^\nu]_\tau
	= \alpha \, \lambda\, g_{\rm E}^{\mu\nu}(Z_\tau) \, d\tau,
\end{equation}
while the second order part of the vector field $w^{^\mu_{\nu\rho}}$ is constructed using the non-rotated metric  \cite{Dohrn:1985iu,Kuipers:2021aok}, such that
\begin{equation}\label{eq:quadVarCond}
	w^{\nu\rho}(Z_\tau,\tau) 
	=
	\alpha \, \lambda \, g^{\nu\rho}(Z_\tau).
\end{equation}
Therefore, eq.~\eqref{eq:VelProc2} only holds under a notationally suppressed transformation as described above.
We also note that $w^{\nu\rho}$ is the second order part of a second order vector field, while $g^{\mu\nu}$ is a bilinear first order tensor. Therefore, the two cannot be equated straightforwardly. However, there exists a unique smooth and invertible linear map $\mathbf{H}$ from bilinear first order forms to second order forms, cf. Proposition 6.13 in Ref.~\cite{Emery}. Using this mapping, which is notationally suppressed in equation \eqref{eq:quadVarCond}, one can equate the two objects.

\section{Variational Equations}
Having specified the geometry and the stochastic dynamics, we can derive equations of motion for the stochastic particle. For this, we assume the geometry to be non-dynamical, and thus the metric to be a fixed symmetric bilinear form $g_{\mu\nu}$. Consequently, the processes $(Z_\tau^\mu,W_\tau^\nu,W_\tau^{\rho\sigma})$ defined on the $\frac{n(n+5)}{2}$-dimensional second order holomorphic tangent bundle $T_2^{1,0}\M$ can be restricted to $(Z_\tau^\mu,W_\tau^\nu)$ defined on the $2n$-dimensional slice $T^{1,0}\M \subset T_2^{1,0}\M$. The Lagrangian for these processes is a complex function on the holomorphic tangent bundle, i.e.,
\begin{align}
	L: T^{1,0}\M \rightarrow \mathbb{C},
\end{align}
and the action is given by
\begin{align}
	S = \E \left[\int L(Z,W) \, d\tau \right].
\end{align}
By considering a variation with respect to a stochastically independent process $(\delta Z,\delta W)$ with $\delta W \, d\tau = \circ\, d\delta Z$, and using that the Stratonovich integral satisfies the usual Leibniz rule, one finds the stochastic Euler-Lagrange equations
\begin{align}
	\int \frac{\p}{\p Z^\mu} L(Z,W) \, d\tau
	&= 
	\dashint \circ \, d \frac{\p}{\p W^\mu} L(Z,W),
\end{align}
which is a set of stochastic differential equation in the sense of Stratonovich.
One can also construct a stochastic Hamiltonian function
\begin{equation}
	H(Z,P) = P_\mu W^\mu - L(Z,W), 
\end{equation}
where $P_\mu$ is the conjugate momentum process, i.e.,
\begin{equation}
	P_\mu = \frac{\p}{\p W^\mu}  L(Z,W).
\end{equation}
\par 

In addition, we define Hamilton's principal function by
\begin{equation}
	S(z,\tau) = \E\left[\int_{0}^{\tau} L(Z,W) \, ds \, \Big|\,  Z_\tau = z  \right].
\end{equation}
The corresponding stochastic Hamilton-Jacobi equations are given by
\begin{align}
	\nabla_\mu S(z,\tau) &= \E_\tau\left[ P_\mu \right],\\
	\frac{\p}{\p \tau} S(z,\tau) &= \E_\tau\left[- H(Z,P) \right].
\end{align}
Finally, we remark that our relativistic theory is invariant under rescalings of the proper time parameter, which imposes
\begin{equation}
	\frac{\p}{\p \tau} S(z,\tau) = 0.
\end{equation}

\section{Stochastic Euler-Lagrange Equations}\label{sec:Lagrangian}
We consider a classical real Lagrangian $L:T\M\rightarrow \R$ of the form
\begin{equation}\label{eq:ClassLag}
	L(x,v) = \frac{1}{2 \lambda} g_{\mu\nu}(x)\, v^\mu v^\nu - \frac{\lambda \, m^2}{2} + q\, A_\mu(x)\, v^\mu,
\end{equation}
where $\lambda$ is an auxiliary variable that fixes the energy-momentum relation. For massive theories we gauge fix $\lambda=m^{-1}$, while for massless theories we gauge fix $\lambda=1$ in the equations of motion.
We consider the stochastic analytic continuation of this Lagrangian given by $L:T^{1,0}\M\rightarrow \mathbb{C}$ such that
\begin{equation}
	L(Z,W) 
	=
	\frac{1}{2\lambda} g_{\mu\nu}(Z)\, W^\mu W^\nu - \frac{\lambda \, m^2}{2}
	+ q\, A_\mu(Z)\, W^\mu.
\end{equation}
The Euler-Lagrange equations for this Lagrangian become
\begin{equation}
	\dashint \dashint g_{\mu\nu} \circ \left(d^2 Z^\nu + \Gamma^\nu_{\rho\sigma} dZ^\rho dZ^\sigma \right) 
	=
	\int \dashint \lambda \, q \left(\nabla_\mu A_\nu - \nabla_\nu A_\mu \right) \circ dZ^\nu d\tau,
\end{equation}
which is a complex second order stochastic differential equation in the sense of Stratonovich. This equation must be supplemented with the relativistic constraint equation
\begin{equation}\label{eq:LineElement}
	\E_{\tau} \left[ g_{\mu\nu} \circ dZ^\mu dZ^\nu + \lambda^2 \, m^2\, d\tau^2\right] = 0
\end{equation}
that follows from the variation of the action with respect to $\lambda$. In addition, it must be supplemented with the condition on the quadratic variation
\begin{equation}
	d[Z^\mu,Z^\nu]
	= \alpha \, \lambda\, g_{\rm E}^{\mu\nu}(Z) \, d\tau.
\end{equation}
\par

We note that in the limit $\alpha\rightarrow0$, one obtains the classical results: the Euler-Lagrange equations become ordinary differential equations
\begin{equation}
	g_{\mu\nu} \left(\frac{d^2 Z^\nu}{d\tau^2} + \Gamma^\nu_{\rho\sigma} \frac{dZ^\rho}{d\tau} \frac{dZ^\sigma}{d\tau} \right) 
	=
	\lambda \, q \, \Big(\nabla_\mu A_\nu - \nabla_\nu A_\mu \Big)\, \frac{dZ^\nu}{d\tau}
\end{equation}
with constraint
\begin{equation}
	g_{\mu\nu}\, \frac{dZ^\mu}{d\tau} \frac{dZ^\nu}{d\tau} = - \lambda^2 \, m^2
\end{equation}
and the quadratic variation vanishes.
\section{Field Equations}
Although the equations of motion derived in the previous section can be written down formally, for practical purposes it may be easier to solve a system of first order stochastic differential equations in the sense of It\^o. In this section, we will therefore derive a system of stochastic differential equations in the It\^o formulation using the Hamilton-Jacobi formalism.
\par 

The Hamilton-Jacobi equations for the Lagrangian introduced in previous section yield
\begin{equation}\label{eq:HamJac1}
	\nabla_\mu S(z,\tau) 
	=
	\lambda^{-1}\, g_{\mu\nu} \hat{w}^\nu + q\, A_\mu.
\end{equation}
and
\begin{align}
	\frac{\p}{\p \tau} S(z,\tau)
	&=
	- \E_{\tau} \left[ \frac{1}{2\lambda}  g_{\mu\nu}(Z)\, W^\mu W^\nu + \frac{\lambda\, m^2}{2} \right]\nonumber\\
	&=
	- \frac{1}{2\lambda} g_{\mu\nu} \hat{w}^\mu \hat{w}^\nu
	- \frac{\alpha}{2} \nabla_\mu \hat{w}^\mu
	+ \frac{\alpha^2 \lambda}{12}\, \mathcal{R}
	- \frac{\lambda\, m^2}{2},\label{eq:HamJac2}
\end{align}
where we used the results from appendix \ref{Ap:CondExp}.
We can combine these two equations by taking a covariant derivative of the second equation and plugging in the first equation. This yields
\begin{equation}
	\frac{1}{\lambda} \, \hat{w}^\nu \, \nabla_\mu \hat{w}_\nu
	+ \frac{\alpha}{2} \nabla_\mu \nabla_\nu \hat{w}^\nu
	- \frac{\alpha^2 \lambda}{12}\, \nabla_\mu \mathcal{R}
	=
	0,
\end{equation}
where we applied the relativistic constraint $\p_\tau S=0$.
Then using that
\begin{align}
	\nabla_\mu \hat{w}_\nu
	&= 
	\nabla_\nu \hat{w}_\mu
	- \lambda \, q \, H_{\mu\nu},\\
	\nabla_\mu \nabla_\nu \hat{w}^\nu
	&= 
	\Box \, \hat{w}_\mu
	- \lambda \, q \, \nabla^\nu H_{\mu\nu}
	- \mathcal{R}_{\mu\nu} \hat{w}^\nu
\end{align}
with the field strength defined by
\begin{equation}
	H_{\mu\nu} := \nabla_\mu A_\nu - \nabla_\nu A_\mu,
\end{equation}
we find
\begin{equation}\label{eq:FieldEq}
	\left[ 
	\frac{1}{\lambda} \, g_{\mu\nu} \, \hat{w}^\rho \, \nabla_\rho
	- q \, H_{\mu\nu}
	+ \frac{\alpha}{2} \Big( 
	g_{\mu\nu}\, \Box
	- \mathcal{R}_{\mu\nu}
	\Big)
	\right] \hat{w}^\nu
	=
	\frac{\alpha \, \lambda}{2} \left(
	q \, \nabla^\nu H_{\mu\nu}
	+ \frac{\alpha}{6} \, \nabla_\mu \mathcal{R}
	\right),
\end{equation}
which can be solved for the velocity field $\hat{w}^\mu(z)$ under the relativistic constraint
\begin{equation}
	g_{\mu\nu} \hat{w}^\mu \hat{w}^\nu
	+ \alpha \, \lambda \nabla_\mu \hat{w}^\mu
	- \frac{\alpha^2 \lambda^2}{6} \mathcal{R}
	=
	- \lambda^2 \, m^2.
\end{equation}
The solution can then be plugged into the first order stochastic differential equation in the sense of It\^o
\begin{align}
	dZ^\mu_\tau &= w^\mu(Z_\tau) \, d\tau + dM^\mu_\tau,\nonumber\\
	d[Z^\mu,Z^\nu]_\tau &= \alpha \, \lambda \, g_{\rm E}^{\mu\nu}(Z_\tau) \, d\tau,
\end{align}
where we note that $\hat{w}^\mu = w^\mu + \frac{\alpha \, \lambda}{2} \, \Gamma^\mu$. This system can be solved for the appropriate boundary conditions, yielding a stochastic process $Z_\tau$. The moments of this process can be calculated using the characteristic and moment generating functional
\begin{align}
	\Phi_{Z}(J) 
	&= \E\left[ e^{{\rm i} \int J_\mu Z^\mu d\tau }\right],\\
	M_{Z}(J)
	&= \E\left[ e^{ \int J_\mu Z^\mu \, d\tau } \right].
\end{align}
\section{Diffusion Equation}
In this section, we derive a diffusion equation governing the stochastic process described in previous sections.
\par

The Hamilton-Jacobi equations \eqref{eq:HamJac1} and \eqref{eq:HamJac2} can be combined such that
\begin{align}\label{DiffS}
	\frac{\p}{\p \tau} S
	&=
	- \frac{\lambda}{2} \left(
	\nabla_\mu S \, \nabla^\mu S
	+ \alpha \, \Box S
	- 2\, q \, A_\mu \, \nabla^\mu S
	- \alpha \, q \, \nabla_\mu  A^\mu
	+ q^2 A_\mu A^\mu 
	- \frac{\alpha^2}{6} \mathcal{R}
	+ m^2
	\right).
\end{align}
If we then define the wave function
\begin{equation}
	\Psi(z,\tau) = \exp\left\{ \frac{1}{\alpha} \left[ S(z,\tau) + \frac{\lambda\, m^2}{2}\, \tau \right] \right\},
\end{equation}
we find that eq.~\eqref{DiffS} is equivalent to the diffusion equation
\begin{equation}\label{eq:Diffusion}
	\frac{\p}{\p \tau} \Psi
	=
	- \frac{\alpha \, \lambda}{2} \left[
	\left( \nabla_\mu - \frac{q}{\alpha} \, A_\mu \right)
	\left( \nabla^\mu - \frac{q}{\alpha} \, A^\mu \right)	
	- \frac{1}{6} \mathcal{R}
	\right] \Psi.
\end{equation}
Moreover, we have
\begin{equation}
	\big|\Psi(z,\tau)\big|^2 
	= 
	\exp\left[ \frac{2}{\rho} \left( 
	\cos(\phi) \left\{ {\rm Re}\big[S(z,\tau)\big] + \frac{\lambda \, m^2}{2} \tau  \right\}
	+ \sin(\phi)\, {\rm Im}\big[S(z,\tau)\big] \right) \right].
\end{equation}
We note that this equation should be interpreted as a backward equation, i.e., subjected to a terminal condition.
\par 

We will now set $\rho=|\alpha|=1$ and consider several special cases. As anticipated in section \ref{sec:Complexification} , for $\phi\in\{0,\pi\}$ we obtain the heat equation
\begin{equation}
	\frac{\p}{\p \tau} \Psi
	=
	\mp \frac{\lambda}{2} \left[
	\Big( \nabla_\mu \mp q \, A_\mu \Big)
	\Big( \nabla^\mu \mp q \, A^\mu \Big)	
	- \frac{1}{6} \mathcal{R}
	\right] \Psi
\end{equation}
with 
\begin{align}
	\Psi(z,\tau) 
	&= 
	\exp\left\{\pm \left[ S(z,\tau)  + \frac{\lambda \, m^2}{2} \tau \right] \right\},\\
	\big|\Psi(z,\tau)\big|^2 
	&= 
	\exp\Big( \pm \Big\{ 2 \, {\rm Re}\big[S(z,\tau)\big] + \lambda \, m^2\, \tau \Big\} \Big).
\end{align}
\par

On the other hand for $\phi\in\{-\frac{\pi}{2},\frac{\pi}{2} \}$, we obtain the Schr\"odinger equation
\begin{equation}
	i \, \frac{\p}{\p \tau} \Psi
	=
	\mp \frac{\lambda}{2} \left[
	\Big( \nabla_\mu \mp i \, q \, A_\mu \Big)
	\Big( \nabla^\mu \mp i \, q \, A^\mu \Big)	
	- \frac{1}{6} \mathcal{R}
	\right] \Psi
\end{equation}
with 
\begin{align}
	\Psi(z,\tau) &= \exp\left\{ \pm\, i \left[ S(z,\tau) + \frac{\lambda \, m^2}{2} \tau \right] \right\},\\
	\big|\Psi(z,\tau)\big|^2 &= \exp\Big\{ \mp 2\, {\rm Im}\big[S(z,\tau)\big] \Big\}.
\end{align}
\par 

Furthermore, we note that the relativistic constraint imposes $S(z,\tau)=S(z)$, which allows to solve eq.~\eqref{eq:Diffusion} by separation of variables. We then obtain
\begin{equation}
	\Psi(z,\tau) = \Phi(z)\, \exp\left( \frac{m \, \tau}{2 \, \alpha} \right),	
\end{equation}
where $\Phi(z)= \exp\left[\alpha^{-1} S(z)\right]$ solves the Klein-Gordon equation
\begin{equation}
	\left[ 
	\left(\nabla_\mu - \frac{q}{\alpha} \, A_\mu \right)
	\left(\nabla^\mu - \frac{q}{\alpha} \, A^\mu \right)	
	- \frac{1}{6} \mathcal{R}
	+ \frac{m^2}{\alpha^2}
	\right]	\Phi
	=
	0.
\end{equation}

\section{Conclusion}
In this paper we have derived three equivalent descriptions for the diffusion of a single scalar relativistic particle on a complexified Lorentzian manifold charged under a vector potential. The first is as a second order stochastic differential equation in the sense of Stratonovich; the second is a system of first order stochastic differential equations in the sense of It\^o and the third is as the Kolmogorov backward equation associated to the process. In addition, we have presented the results for the non-relativistic particle in appendix \ref{Ap:NonRel}.
\par 

In fact, this result is well known for non-relativistic diffusion processes on $\R^n$ with a real quadratic variation, and is given by the Feynman-Kac formula. In this paper, we have used Nelson's stochastic quantization procedure to generalize this result to the case of (relativistic) diffusion processes on (pseudo-)Riemannian manifolds with a complex quadratic variation. We should emphasize, however, that we have derived our results under the assumption of the existence of unique solutions to the given formulations. A mathematically rigorous proof of our results will be left for future work. Moreover, we point out that, in contrast to stochastic mechanics, we have not derived the Born rule. Instead, the stochastic interpretation of the wave function is provided by the generalization of the Feynman-Kac formula.
\par 

It is worth pointing out the similarities and differences of the rotation of the complex quadratic variation around the angle $\phi$ studied in this paper and the Wick rotation. Both rotations transform a heat-type equation into a Schr\"odinger-type equation. This is due to the fact that both rotations act on the proper time parameter. However, there is also an important difference, as the rotation discussed in this paper also acts on all coordinates. As a consequence it preserves the $(k,l,m)$ signature of the (pseudo-)Riemannian manifold. In contrast the Wick rotation only acts on the time-like coordinates, and therefore transforms a pseudo-Riemannian manifold with $(k,l,m)$ signature into a Riemannian manifold with $(k+l,0,m)$ signature.
\par

Furthermore, it is worth noticing that the diffusion equation \eqref{eq:Diffusion} contains a term proportional to the Ricci scalar. This term comes with a prefactor $\frac{1}{6}$ that results from a Taylor expansion, cf. appendix \ref{Ap:CondExp}. On the other hand, it is well known that for a prefactor given by $\frac{n-2}{4(n-1)}$ the diffusion equation is conformally invariant for $m=0$. Interestingly, the two prefactors coincide in $4$ dimensions.
\par

Finally, as the description given in this paper requires the complexification of spacetime, we were forced to give a physical interpretation to the the notion of an imaginary position. We then gave the interpretation that ${\rm Re}(W)=V$ is the velocity of matter, while ${\rm Im}(W)=U$ is the velocity of the spacetime foam. Consequently, we interpreted ${\rm Re}(Z)=X$ as the position of the particle and ${\rm Im}(Z)=Y$ as the position of an associated particle in the spacetime foam. We concluded that both the spacetime itself and the matter defined on this spacetime move under evolution of the proper time. Interestingly, for $\alpha\in\mathbb{C}\setminus\R$ the stochastic dynamics of the particle and the spacetime foam are coupled, and, therefore, they cannot be treated independently. This is particularly true for pure quantum systems where $\alpha\in {\rm i}\, \R$, but is in stark contrast with the real Brownian motion where $\alpha\in\R$. In this latter case, the motion of the spacetime foam and matter are completely decoupled, which allows to neglect the motion of the spacetime foam.
\par 

We conclude that our results further illustrate the close connection between Brownian motion and quantum physics and open up new avenues to tackle quantum problems using the theory of stochastic differential equations. In addition, our results reaffirm the central result of stochastic mechanics that quantum physics can be understood in terms of stochastic processes. Finally, our results hint towards a possible unification of statistical physics and quantum physics in a larger framework of complex stochastic physics.

\section*{Acknowledgments}
This work is supported by a doctoral studentship of the Science and Technology Facilities Council, award number: 2131791.

\appendix

\section{Non-Relativistic Theories}\label{Ap:NonRel}
In this paper, we have presented a stochastic formulation of relativistic diffusion processes. In this appendix we present the results for non-relativistic diffusion processes, which can be derived in a similar fashion.
\par

We consider a set $\mathcal{T}=[0,T]$, a real $(n=d)$-dimensional Riemannian manifold and trajectories $x(t):\mathcal{T}\rightarrow\M$. We consider a classical non-relativistic theory of the form
\begin{equation}
	L(x,v,t) = \frac{m}{2} \, g_{ij}(x) \, v^i v^j + q \, A_i(x,t) \, v^i - \mathfrak{U}(x,t).
\end{equation}
The stochastic analytic continuation is then given by
\begin{equation}
	L(Z,W,t) = \frac{m}{2} \, g_{ij}(Z) \, W^i W^j + q \, A_i(Z,t) \, W^i - \mathfrak{U}(Z,t)
\end{equation}
and the stochastic Euler-Lagrange equations are
\begin{equation}
	m \, g_{ij} \circ \left(d^2 Z^j + \Gamma^j_{kl}\, dZ^k dZ^l \right) 
	=
	q \, \Big(\nabla_i A_j - \nabla_j A_i \Big)\circ dZ^j dt
	- \Big( q \, \p_t A_i
	+ \nabla_i \mathfrak{U} \Big) \, dt^2,
\end{equation}
which must be supplemented with the condition on the quadratic variation
\begin{equation}
	d[Z^i, Z^j] = \frac{\alpha}{m} \, g^{ij}(Z) \, dt.
\end{equation}
\par

On te other hand, in the It\^o formulation, we find that the velocity field is governed by the equation
\begin{align}
	\left[
	m \, g_{ij} \, \Big(
	\p_t 
	+ \hat{w}^k \nabla_k \Big)
	- q \, H_{ij}
	+ \frac{\alpha}{2} \Big( 
	g_{ij}\, \Box
	- \mathcal{R}_{ij}
	\Big)
	\right] \hat{w}^j
	&=
	\frac{\alpha \, q}{2 m}\, \nabla^j H_{ij}
	- q \, \p_t A_i
	- \nabla_i \mathfrak{U}
	+ \frac{\alpha^2}{12 m} \, \nabla_i \mathcal{R}.
\end{align}
As in the relativistic case the solution $w^i(z,t)$ can be plugged into the first order stochastic differential equation in the sense of It\^o:
\begin{align}
	dZ^i_t &= w^i(Z_t,t) \, dt + dM^i_t,\nonumber\\
	d[Z^i,Z^j]_t &= \frac{\alpha}{m} g^{ij}(Z_t) \, dt,
\end{align}
where we note that $\hat{w}^i = w^i + \frac{\alpha}{2 m} \Gamma^i$.
\par 

Furthermore, we can define the wave function
\begin{equation}
	\Psi(z,t)
	= 
	\exp \left[ \frac{S(z,t)}{\alpha}  \right]\, ,
\end{equation}
which satisfies the complex diffusion equation
\begin{equation}
	\alpha \, \frac{\p}{\p t} \Psi
	=
	- \left\{ 
	\frac{\alpha^2}{2 m} \left[
	\Big(\nabla_i - \frac{q}{\alpha} \, A_i \Big)
	\Big(\nabla^i - \frac{q}{\alpha} \, A^i \Big)	
	- \frac{1}{6} \mathcal{R}
	\right]
	+ \mathfrak{U}
	\right\} \Psi.
\end{equation}
If there is no explicit time dependence, i.e. $\mathfrak{U}(x,t)=\mathfrak{U}(x)$ and $A_i(x,t)=A_i(x)$, this can be solved by separation of variables, such that
\begin{equation}
	\Psi(z,t) = \sum_k \Phi_k(z) \, \exp\left[\frac{E_k}{\alpha} \, t \right],
\end{equation}
where $\Phi_k(z)$ solves the wave equation
\begin{equation}
	\left\{ 
	\frac{\alpha^2}{2 m} \left[
	\Big(\nabla_i - \frac{q}{\alpha} \, A_i \Big)
	\Big(\nabla^i - \frac{q}{\alpha} \, A^i \Big)	
	- \frac{1}{6} \mathcal{R}
	\right]
	+ \mathfrak{U}
	+ E_k
	\right\} \Psi_k
	=
	0.
\end{equation}

\section{Stochastic Integration}\label{Ap:StochInt}
In this appendix, we review some notions from stochastic integration on manifolds. Let us first review the definition of stochastic integrals on $\R^n$. The \textit{Stratonovich integral} is defined as 
\begin{equation}\label{eq:Stratonovich}
	\dashint_0^T f(X_\tau) \circ dX^\mu_\tau
	:=
	\lim_{k\rightarrow\infty} \sum_{[\tau_i,\tau_{i+1}]\in\pi_k} 
	\frac{1}{2} \big[ f(X_{\tau_i}) + f(X_{\tau_{i+1}}) \big] \big[X_{\tau_{i+1}}^\mu - X_{\tau_i}^\mu \big],
\end{equation}
where $\pi_k$ is a partition of $[0,T]$. The \textit{It\^o integral} is defined by
\begin{equation}\label{eq:Ito}
	\lowint_0^T f(X_\tau) \, dX^\mu_\tau
	:=
	\lim_{k\rightarrow\infty} \sum_{[\tau_i,\tau_{i+1}]\in\pi_k} 
	f(X_{\tau_i}) \, \big[X_{\tau_{i+1}}^\mu - X_{\tau_i}^\mu \big]
\end{equation}
and the integral over the quadratic variation is given by
\begin{equation}\label{eq:QuadVar}
	\int f(X_\tau) \, d[X^\mu,X^\nu]_\tau
	:=
	\lim_{k\rightarrow\infty} \sum_{[\tau_i,\tau_{i+1}]\in\pi_k} 
	f(X_{\tau_i}) \, \big[X_{\tau_{i+1}}^\mu - X_{\tau_i}^\mu \big] \big[X_{\tau_{i+1}}^\nu - X_{\tau_i}^\nu \big]\, .
\end{equation}
By a straightforward calculation, one can then derive a relation between the three integrals:
\begin{equation}
	\dashint_0^T f(X_\tau) \, dX^\mu_\tau
	=
	\lowint_0^T f(X_\tau) \, dX^\mu_\tau
	+ \frac{1}{2} \int \p_\nu f(X_\tau) \, d[X^\mu,X^\nu]_\tau
\end{equation}
The Stratonovich integral has the advantage that it obeys the Leibniz rule:
\begin{equation}
	\circ \, d(X^\mu Y^\nu) = X^\mu\circ dY^\nu + Y^\nu\circ dX^\mu,
\end{equation}
while the It\^o integral satisfies a modified Leibniz rule given by
\begin{equation}
	d(X^\mu Y^\nu) = X^\mu\, dY^\nu + Y^\nu dX^\mu + d[X^\mu,Y^\nu].
\end{equation}
On the other hand, the It\^o integral has the advantage that for any martingale $M_\tau$
\begin{equation}\label{eq:MartProp}
	\E_\tau \left[\int_{\tau}^T f(X_s) \, dM_s^\mu \right] = 0.
\end{equation}
\par 

All these integrals can be extended to smooth manifolds with a connection. As usual this must be done using differential forms. We will express a first order form $\omega\in T^\ast\M$ in a local coordinate chart as
\begin{equation}
	\omega = \omega_\mu \circ dx^\mu.
\end{equation}
The Stratonovich integral is then defined by
\begin{equation}
	\dashint_{X_\tau} \omega := \dashint_0^T \omega_\mu(X_\tau) \circ dX^\mu_\tau.
\end{equation}
The right hand side can be calculated using the definition \eqref{eq:Stratonovich} in a local coordinate chart.
\par 

The construction of the It\^o integral on the other hand, requires the construction of second order forms $\Omega\in T^\ast_2\M$. These can be expressed in a local coordinate chart as\footnote{Note that we deviate here from the notation used in Refs.~\cite{Emery,Kuipers:2021jlh}, where first order forms are expressed as $\omega=\omega_\mu dx^\mu$ and second order forms as $\omega=\omega_\mu d_2x^\mu + \omega_{\mu\nu} \, dx^\mu \cdot dx^\nu$. The notation used in Refs.~\cite{Emery,Kuipers:2021jlh} is the standard notation in the geometry literature, while the notation adapted in this paper is closer to the stochastics literature.}
\begin{equation}
	\omega = \omega_\mu \, d x^\mu + \frac{1}{2}\, \p_\nu \omega_\mu \, d[x^\mu,x^\nu]
\end{equation}
Expressions of the form
\begin{equation*}
	\lowint_0^T \omega_\mu(X_\tau) \, d X_\tau^\mu
	\qquad {\rm and} \qquad
	\int_0^T \omega_{\mu\nu}(X_\tau) \, d[X^\mu,X^\nu]_\tau
\end{equation*}
can then be calculated in a local coordinate chart using definitions \eqref{eq:Ito} and \eqref{eq:QuadVar} respectively. Moreover, the second expression represents the integral over the quadratic variation on a manifold. The first, however, does not define an It\^o integral on manifolds, as it is not covariant. Instead, the It\^o integral is defined by the covariant expression
\begin{align}
	\lowint_{X_\tau} \omega 
	&:= 
	\lowint_0^T \omega_\mu(X_\tau) \, d \hat{X}_\tau^\mu  \nonumber\\
	&:=
	\lowint_0^T \omega_\mu(X_\tau) \, d X_\tau^\mu 
	+ \frac{1}{2} \int_0^T \omega_{\mu}(X_\tau) \, \Gamma^\mu_{\nu\rho}(X_\tau) \, d[X^\nu,X^\rho]_\tau.
\end{align}
The relation between the Stratonovich and It\^o integral on a manifold is then given by
\begin{align}
	\dashint_0^T \omega_\mu(X_\tau) \circ dX^\mu_\tau
	= 
	\lowint_{X_\tau} \omega_\mu(X_\tau) \, d\hat{X}^\mu_\tau
	+ \frac{1}{2} \int_0^T \nabla_\nu \omega_\mu(X_\tau) \, d[X^\mu,X^\nu]_\tau.
\end{align}

\section{Calculation of conditional expectations}\label{Ap:CondExp}
In this appendix we derive the following expressions
\begin{align}
	\E_\tau \big[ \mathfrak{U} \big] 
	&= 
	\mathfrak{U},\\
	\E_\tau \big[ g_{\mu\nu} W^{\mu\nu} \big] 
	&= 
	n \, \alpha \, \lambda, \label{eq:CondQuadVar}\\
	\E_\tau \big[ A_\mu W^\mu \big] 
	&= 
	A_\mu \hat{w}^\mu 
	+ \frac{\alpha \, \lambda}{2} \, \nabla_\mu A^\mu,\label{eq:CondExpLin}\\
	\E_\tau\big[ g_{\mu\nu}\, W^\mu W^\nu \big]
	&=
	g_{\mu\nu} \hat{w}^\mu \hat{w}^\nu 
	+ \alpha \, \lambda \, \nabla_\mu \hat{w}^\mu
	- \frac{\alpha^2\lambda^2}{6} \, \mathcal{R}.\label{eq:CondExpQuad}
\end{align}
\par 

The proof of the first equality is immediate by ``taking out what is known'':
\begin{equation}
	\E_\tau\left[ \mathfrak{U}(Z_\tau) \right]
	= \mathfrak{U}(z).
\end{equation}
\par

For the second equality we find
\begin{align}\label{eq:CondExpDerQvar}
	\E_\tau \left[\int_{\tau}^{\tau+d\tau} g_{\mu\nu} (Z_s)\, W_s^{\mu\nu} \, ds \right]
	&=
	\E_\tau \left[ \int g_{\mu\nu} (Z_s) \, d[Z^\mu,Z^\nu]_s \right] \nonumber\\
	&=
	\E_\tau \left[ \alpha \, \lambda \int g_{\mu\nu}(Z_s) \, g^{\mu\nu}(Z_s)\, ds \right]\nonumber\\
	&=
	\E_\tau \big[ n \, \alpha \, \lambda  \, d\tau \big]\nonumber\\
	&=
	n \, \alpha \, \lambda \, d\tau,
\end{align}
In the limit $d\tau\rightarrow0$ we then obtain the result \eqref{eq:CondQuadVar}.
\par

For the third equality, we find
\begin{align}\label{eq:CondExpDer}
	\E_\tau \left[\int_{\tau}^{\tau+d\tau} A_\mu (Z_s)\, W_s^\mu \, ds \right]
	&=
	\E_\tau \left[ \dashint A_\mu(Z_s) \circ d Z_s^\mu \right] \nonumber\\
	&=
	\E_\tau \left[ \lowint A_\mu (Z_s)\, d Z_s^\mu
	+ \frac{1}{2} \int \p_\nu A_\mu(Z_s)\, d[Z^\mu,Z^\nu]_s \right] \nonumber\\
	&=
	\E_\tau \left[ \int \Big( A_\mu(Z_s)\, w^\mu(Z_s) + w^{\mu\nu}(Z_s)\, \p_\nu A_\mu(Z_s) \Big) ds 
	+ \lowint A_\mu (Z_s)\, d M_s^\mu \right] \nonumber\\
	&=
	\E_\tau \left[ \Big( A_\mu(Z_\tau)\, w^\mu(Z_\tau) + w^{\mu\nu}(Z_\tau) \, \p_\nu A_\mu(Z_\tau) \Big) d\tau + o(d\tau) \right] \nonumber\\
	&=
	\E_\tau \left[ \Big( A_\mu(Z_\tau)\, \hat{w}^\mu(Z_\tau) + \hat{w}^{\mu\nu}(Z_\tau) \, \nabla_\nu A_\mu(Z_\tau) \Big) d\tau + o(d\tau) \right] \nonumber\\
	&= 
	\left( A_\mu \, \hat{w}^\mu  + \frac{\alpha\, \lambda}{2} \, \nabla_\mu A^\mu \right) d\tau + o(d\tau),
\end{align}
where we rewrote the Stratonovich integral as an It\^o integral, such that the martingale property \eqref{eq:MartProp} can be applied on the stochastic integral $dM$. In the limit $d\tau\rightarrow 0$, we then obtain eq.~\eqref{eq:CondExpLin}.

\subsection{Quadratic in Velocity}
The calculation of the conditional expectation of a term quadratic in the velocity process is slightly more involved. This calculation was first performed by Guerra and Nelson in Ref.~\cite{Nelson}. Here, we reproduce their result using a slightly different presentation.
\par

We first notice that
\begin{equation}\label{eq:QuadVelProc}
	g_{\mu\nu}(Z_\tau) \circ dZ_\tau^\mu\, dZ_\tau^\nu
	=
	g_{\mu\nu}(Z_\tau)\, W_\tau^{\mu\nu} \, d\tau
	+ g_{\mu\nu}(Z_\tau)\, W_\tau^\mu W_\tau^\nu \, d\tau^2
	+ o(d\tau^2),
\end{equation}
where the left hand side is a Stratonovich integral. In order to calculate the conditional expectation of this expression, we will need to rewrite this into an It\^o integral. For this, we note that\footnote{We make use of the that Brownian motion is completely determined by its quadratic moment: all even moment can be expressed in terms of the quadratic moment and all odd moments vanish.}
\begin{align}
	d^2f
	&=
	d\left(
	\p_\mu f \, dZ^\mu 
	+ \frac{1}{2} \p_\mu \p_\nu f \, d[Z^\mu,Z^\nu]  
	\right)
	\nonumber\\
	&=
	\p_\mu f \, d^2Z^\mu
	+ \p_\nu \p_\mu f \, dZ^\mu dZ^\nu
	+ \p_\rho \p_\nu \p_\mu f \, dZ^\mu \, d[Z^\nu,Z^\rho] 
	\nonumber\\
	&\quad
	+ \frac{1}{4} \, \p_\sigma \p_\rho \p_\nu \p_\mu f \,
	d[Z^\mu,Z^\nu] \, d[Z^\rho,Z^\sigma]
	\nonumber\\
	&=
	\p_\mu f \, d^2Z^\mu
	+ \p_\nu \p_\mu f \, dZ^\mu dZ^\nu
	+ \frac{1}{3} \, \p_\rho \p_\nu \p_\mu f \, dZ^\mu dZ^\nu dZ^\rho 
	\nonumber\\
	&\quad
	+ \frac{1}{12} \, \p_\sigma \p_\rho \p_\nu \p_\mu f \,
	dZ^\mu dZ^\nu dZ^\rho dZ^\sigma,
\end{align}
where we introduced the notation
\begin{align}
	dZ^\mu dZ^\nu dZ^\rho 
	&=
	dZ^\mu \, d[Z^\nu,Z^\rho] + dZ^\nu \, d[Z^\mu,Z^\rho] + dZ^\rho \, d[Z^\mu,Z^\nu],\\
	dZ^\mu dZ^\nu dZ^\rho dZ^\sigma 
	&= 
	d[Z^\mu,Z^\nu] \, d[Z^\rho,Z^\sigma] + d[Z^\mu,Z^\rho] \, d[Z^\nu,Z^\sigma]
	+ d[Z^\mu,Z^\sigma] \, d[Z^\nu,Z^\rho].
\end{align}
This expression can be rewritten into an explicitly covariant form:
\begin{align}
	d^2 f 
	&=
	\nabla_\mu f \left[
	dZ^\mu
	+ \Gamma^\mu_{\nu\rho} \, dZ^\nu dZ^\rho
	+ \frac{1}{3} \Big( \p_\nu \Gamma^\mu_{\rho\sigma} + \Gamma^\mu_{\nu\kappa} \Gamma^\kappa_{\rho\sigma} \Big) dZ^\nu dZ^\rho dZ^\sigma 
	\right. \nonumber\\
	&\qquad
	+ \frac{1}{12} \, \p_\kappa \Big( \p_\nu \Gamma^\mu_{\rho\sigma} + \Gamma^\mu_{\nu\lambda} \Gamma^\lambda_{\rho\sigma}  \Big) dZ^\nu dZ^\rho dZ^\sigma dZ^\kappa
	\nonumber\\
	&\qquad \left.
	+ \frac{1}{12} \, \Gamma^\mu_{\kappa\lambda} \Big( \p_\nu \Gamma^\lambda_{\rho\sigma} + \Gamma^\lambda_{\nu\alpha} \Gamma^\alpha_{\rho\sigma}  \Big) dZ^\nu dZ^\rho dZ^\sigma dZ^\kappa
	\right]
	\nonumber\\
	&\quad
	+ \nabla_\nu \nabla_\mu f \left[
	dZ^\mu dZ^\nu
	+ \frac{2}{3} \, \Gamma^\mu_{\rho\sigma} \, dZ^\nu dZ^\rho dZ^\sigma
	+ \frac{1}{3} \, \Gamma^\nu_{\rho\sigma} \, dZ^\mu dZ^\rho dZ^\sigma
	\right. \nonumber\\
	&\qquad 
	+ \frac{1}{4} \, \Gamma^\mu_{\rho\sigma} \Gamma^\nu_{\kappa\lambda} \, dZ^\rho dZ^\sigma dZ^\kappa dZ^\lambda
	+ \frac{1}{4} \Big( \p_\kappa \Gamma^\mu_{\rho\sigma} + \Gamma^\mu_{\kappa\lambda} \Gamma^\lambda_{\rho\sigma} \Big) dZ^\nu dZ^\rho dZ^\sigma dZ^\kappa
	\nonumber\\
	&\qquad \left.
	+ \frac{1}{12} \Big( \p_\kappa \Gamma^\nu_{\rho\sigma} + \Gamma^\nu_{\kappa\lambda} \Gamma^\lambda_{\rho\sigma} \Big) dZ^\mu dZ^\rho dZ^\sigma dZ^\kappa
	\right]
	\nonumber\\
	&\quad
	+ \frac{1}{3} \, \nabla_\rho \nabla_\nu \nabla_\mu f \left(
	dZ^\mu dZ^\nu dZ^\rho
	+ \frac{3}{4} \, \Gamma^\mu_{\sigma\kappa} \, dZ^\nu dZ^\rho dZ^\sigma dZ^\kappa
	\right.\nonumber\\
	&\qquad \left.
	+ \frac{1}{2} \, \Gamma^\nu_{\sigma\kappa} \, dZ^\mu dZ^\rho dZ^\sigma dZ^\kappa
	+ \frac{1}{4} \, \Gamma^\rho_{\sigma\kappa} \, dZ^\mu dZ^\nu dZ^\sigma dZ^\kappa
	\right)
	\nonumber\\
	&\quad
	+ \frac{1}{12} \, \nabla_\sigma \nabla_\rho \nabla_\nu \nabla_\mu f \,
	dZ^\mu dZ^\nu dZ^\rho dZ^\sigma,
\end{align}
and therefore
\begin{align}
	d^2f
	&=
	\nabla_\mu f \left[
	dZ^\mu
	+ \Gamma^\mu_{\nu\rho} \, dZ^\nu dZ^\rho
	+ \frac{1}{3} \Big( \p_\nu \Gamma^\mu_{\rho\sigma} + \Gamma^\mu_{\nu\kappa} \Gamma^\kappa_{\rho\sigma} \Big) dZ^\nu dZ^\rho dZ^\sigma 
	\right. \nonumber\\
	&\qquad
	+ \frac{1}{12} \, \p_\kappa \Big( \p_\nu \Gamma^\mu_{\rho\sigma} + \Gamma^\mu_{\nu\lambda} \Gamma^\lambda_{\rho\sigma}  \Big) dZ^\nu dZ^\rho dZ^\sigma dZ^\kappa
	\nonumber\\
	&\qquad
	+ \frac{1}{12} \, \Gamma^\mu_{\kappa\lambda} \Big( \p_\nu \Gamma^\lambda_{\rho\sigma} + \Gamma^\lambda_{\nu\alpha} \Gamma^\alpha_{\rho\sigma}  \Big) dZ^\nu dZ^\rho dZ^\sigma dZ^\kappa\
	\nonumber\\
	&\qquad \left.
	+ \frac{1}{12} \, \Gamma^\lambda_{\rho\sigma} \mathcal{R}^\mu_{\;\;\nu\lambda\kappa} dZ^\nu dZ^\rho dZ^\sigma dZ^\kappa
	\right]
	\nonumber\\
	&\quad
	+ \nabla_{(\nu} \nabla_{\mu)} f \left[
	dZ^\mu dZ^\nu
	+ \frac{1}{2} \, \Gamma^\mu_{\rho\sigma} \, dZ^\nu dZ^\rho dZ^\sigma
	+ \frac{1}{2} \, \Gamma^\nu_{\rho\sigma} \, dZ^\mu dZ^\rho dZ^\sigma
	\right. \nonumber\\
	&\qquad 
	+ \frac{1}{4} \, \Gamma^\mu_{\rho\sigma} \Gamma^\nu_{\kappa\lambda} \, dZ^\rho dZ^\sigma dZ^\kappa dZ^\lambda
	+ \frac{1}{6} \Big( \p_\kappa \Gamma^\mu_{\rho\sigma} + \Gamma^\mu_{\kappa\lambda} \Gamma^\lambda_{\rho\sigma} \Big) dZ^\nu dZ^\rho dZ^\sigma dZ^\kappa
	\nonumber\\
	&\qquad \left.
	+ \frac{1}{6} \Big( \p_\kappa \Gamma^\nu_{\rho\sigma} + \Gamma^\nu_{\kappa\lambda} \Gamma^\lambda_{\rho\sigma} \Big) dZ^\mu dZ^\rho dZ^\sigma dZ^\kappa
	\right]
	\nonumber\\
	&\quad
	+ \frac{1}{3} \, \nabla_{(\rho} \nabla_\nu \nabla_{\mu)} f \left(
	dZ^\mu dZ^\nu dZ^\rho
	+ \frac{1}{2} \, \Gamma^\mu_{\sigma\kappa} \, dZ^\nu dZ^\rho dZ^\sigma dZ^\kappa
	\right.\nonumber\\
	&\qquad \left.
	+ \frac{1}{2} \, \Gamma^\nu_{\sigma\kappa} \, dZ^\mu dZ^\rho dZ^\sigma dZ^\kappa
	+ \frac{1}{2} \, \Gamma^\rho_{\sigma\kappa} \, dZ^\mu dZ^\nu dZ^\sigma dZ^\kappa
	\right)
	\nonumber\\
	&\quad
	+ \frac{1}{12} \, \nabla_{(\sigma} \nabla_\rho \nabla_\nu \nabla_{\mu)} f \,
	dZ^\mu dZ^\nu dZ^\rho dZ^\sigma.
\end{align}
By reading of the term proportional to $\nabla_{\mu}\nabla_{\nu}f$, we conclude
\begin{align}
	g_{\mu\nu} \circ dZ_\tau^\mu dZ_\tau^\nu
	&=
	g_{\mu\nu} \left[ dZ_\tau^\mu \, dZ_\tau^\nu
	+ \Gamma^\mu_{\rho\sigma} \, dZ_\tau^\nu  dZ_\tau^\rho dZ_\tau^\sigma
	+ \frac{1}{4} \Gamma^\mu_{\rho\sigma} \Gamma^\nu_{\kappa\lambda} \, dZ_\tau^\rho dZ_\tau^\sigma dZ_\tau^\kappa dZ_\tau^\lambda
	\right. \nonumber\\
	&\qquad \quad \left.
	+ \frac{1}{3} \, \Big( 
	\p_\kappa \Gamma^\mu_{\rho\sigma} + \Gamma^\mu_{\kappa\lambda} \Gamma^\lambda_{\rho\sigma} \Big) \, dZ_\tau^\nu dZ_\tau^\rho dZ_\tau^\sigma dZ_\tau^\kappa \right].
\end{align}
where the It\^o differential is given by
\begin{align}
	dZ_\tau^\mu
	&= Z^\mu_{\tau+d\tau} - Z^\mu_\tau \nonumber\\
	&= \int_{\tau}^{d\tau} w^\mu(Z_s)\, ds + dM_{\tau}^\mu
\end{align}
\par 

We can now calculate the conditional expectation of this expression. We find
\begin{align}
	\E_\tau \Big[ dZ^\mu_\tau \, dZ^\nu_\tau \Big]
	&=
	\E_\tau \left[
	dM_\tau^\mu dM_\tau^\nu
	+ dM_\tau^\mu \int_{\tau}^{\tau+d\tau} w^\nu(Z_s) \, ds
	+ dM_\tau^\nu \int_{\tau}^{\tau+d\tau} w^\mu(Z_s) \, ds
	\right. \nonumber\\
	&\qquad \left.
	+ \int_{\tau}^{\tau+d\tau} w^\mu(Z_s) \, ds \int_{\tau}^{\tau+d\tau} w^\nu(Z_r) \, dr
	+ o(d\tau^2)
	\right] 
	\nonumber\\
	&=
	\E_\tau \left[
	\int_{\tau}^{\tau+d\tau} w^{\mu\nu}(Z_s)\, ds
	+ dM_\tau^\mu \int_{\tau}^{\tau+d\tau} w^\nu(Z_s) \, ds
	+ dM_\tau^\nu \int_{\tau}^{\tau+d\tau} w^\mu(Z_s) \, ds
	\right. \nonumber\\
	&\qquad \left.
	+ \int_{\tau}^{\tau+d\tau} w^\mu(Z_s) \, ds \int_{\tau}^{\tau+d\tau} w^\nu(Z_r) \, dr
	+ o(d\tau^2)
	\right] 
	\nonumber\\
	&=
	\E_\tau \left[
	w^{\mu\nu}(Z_\tau) \int_{\tau}^{\tau+d\tau} ds
	+ \p_\rho w^{\mu\nu}(Z_\tau) \int_{\tau}^{\tau+d\tau} \left( M_s^\rho - M_\tau^\rho \right) \, ds
	\right. \nonumber\\
	&\qquad
	+ \p_\rho w^{\mu\nu}(Z_\tau) \int_{\tau}^{\tau+d\tau} \int_{\tau}^{s} w^\rho(Z_r) \, dr \, ds
	\nonumber\\
	&\qquad
	+ \frac{1}{2} \, \p_\rho \p_\sigma w^{\mu\nu}(Z_\tau) \int_{\tau}^{\tau+d\tau} 
	\left( M_s^\rho - M_\tau^\rho \right)  \left( M_s^\sigma - M_\tau^\sigma \right) \, ds
	\nonumber\\
	&\qquad
	+ w^\nu(Z_\tau) \, dM_\tau^\mu \int_{\tau}^{\tau+d\tau}  ds
	+ \p_\rho w^\nu(Z_\tau) \, dM_\tau^\mu \int_{\tau}^{\tau+d\tau} \left( M_s^\rho - M_\tau^\rho \right)  ds
	\nonumber\\
	&\qquad
	+ w^\mu(Z_\tau) \, dM_\tau^\nu \int_{\tau}^{\tau+d\tau}  ds
	+ \p_\rho w^\mu(Z_\tau) \, dM_\tau^\nu \int_{\tau}^{\tau+d\tau} \left( M_s^\rho - M_\tau^\rho \right)  ds
	\nonumber\\
	&\qquad \left.
	+ w^\mu(Z_\tau) \, w^\nu(Z_\tau) \int_{\tau}^{\tau+d\tau} ds \int_{\tau}^{\tau+d\tau} dr
	+ o(d\tau^2)
	\right] 
	\nonumber\\
	&=
	\E_\tau \Big[
	w^{\mu\nu}(Z_\tau) \, d\tau
	+ w^\rho(Z_\tau) \, \p_\rho w^{\mu\nu}(Z_\tau) \int_{\tau}^{\tau+d\tau} (s-\tau) \, ds
	\nonumber\\
	&\qquad
	+ \frac{1}{2} \, \p_\rho \p_\sigma w^{\mu\nu}(Z_\tau) \int_{\tau}^{\tau+d\tau} \int_{\tau}^s w^{\rho\sigma}(Z_r) \, dr \, ds
	\nonumber\\
	&\qquad
	+ \p_\rho w^\nu(Z_\tau) \int_{\tau}^{\tau+d\tau} \int_\tau^{s} w^{\mu\rho}(Z_r) \, dr \, ds
	+ \p_\rho w^\mu(Z_\tau) \int_{\tau}^{\tau+d\tau} \int_\tau^{s} w^{\nu\rho}(Z_r) \, dr \, ds
	\nonumber\\
	&\qquad
	+ w^\mu(Z_\tau) \, w^\nu(Z_\tau) \, d\tau^2
	+ o(d\tau^2)
	\Big] 
	\nonumber\\
	&=
	w^{\mu\nu}(Z_\tau) \, d\tau
	+ \frac{1}{2} \, w^\rho(Z_\tau) \, \p_\rho w^{\mu\nu}(Z_\tau)\, d\tau^2
	+ \frac{1}{4} \,  w^{\rho\sigma}(Z_\tau) \, \p_\rho \p_\sigma w^{\mu\nu}(Z_\tau) \, d\tau^2
	\nonumber\\
	&\quad
	+ \frac{1}{2} \, w^{\mu\rho}(Z_\tau) \, \p_\rho w^\nu(Z_\tau) \, d\tau^2
	+ \frac{1}{2} \, w^{\nu\rho}(Z_\tau) \, \p_\rho w^\mu(Z_\tau) \, d\tau^2
	\nonumber\\
	&\quad
	+ w^\mu(Z_\tau) \, w^\nu(Z_\tau) \, d\tau^2
	+ o(d\tau^2),
\end{align}
\begin{align}
	\E_\tau \Big[ dZ_\tau^\nu dZ_\tau^\rho dZ_\tau^\sigma \Big]
	&=
	\E_\tau \left[
	dM_\tau^\nu dM_\tau^\rho \int_{\tau}^{\tau+d\tau} w^\sigma(Z_s) \, ds 
	+ dM_\tau^\nu dM_\tau^\sigma \int_{\tau}^{\tau+d\tau} w^\rho(Z_s) \, ds 
	\right. \nonumber\\
	&\qquad \left.
	+ dM_\tau^\rho dM_\tau^\sigma \int_{\tau}^{\tau+d\tau} w^\nu(Z_s) \, ds 
	+ dM_\tau^\nu dM_\tau^\rho dM_\tau^\sigma
	+ o(d\tau^2) \right]
	\nonumber\\
	&=
	\E_\tau \left[
	\int_{\tau}^{\tau+d\tau} w^\nu(Z_s) \, ds \int_{\tau}^{\tau+d\tau} w^{\rho\sigma}(Z_r) \, dr
	+ dM_\tau^\nu \int_{\tau}^{\tau+d\tau} w^{\rho\sigma}(Z_s) \, ds
	\right. \nonumber\\
	&\qquad 
	+ \int_{\tau}^{\tau+d\tau} w^\rho(Z_s) \, ds \int_{\tau}^{\tau+d\tau} w^{\nu\sigma}(Z_r) \, dr
	+ dM_\tau^\rho \int_{\tau}^{\tau+d\tau} w^{\nu\sigma}(Z_s) \, ds
	\nonumber\\
	&\qquad \left.
	+ \int_{\tau}^{\tau+d\tau} w^\sigma(Z_s) \, ds \int_{\tau}^{\tau+d\tau} w^{\nu\rho}(Z_r) \, dr
	+ dM_\tau^\sigma \int_{\tau}^{\tau+d\tau} w^{\nu\rho}(Z_s) \, ds \right]
	+ o(d\tau^2)
	\nonumber\\
	&=
	\E_\tau \Big[
	w^\nu(Z_\tau) \, w^{\rho\sigma}(Z_\tau) \, d\tau^2
	+ w^\rho(Z_\tau) \, w^{\nu\sigma}(Z_\tau) \, d\tau^2
	+ w^\sigma(Z_\tau) \, w^{\nu\rho}(Z_\tau) \, d\tau^2
	\nonumber\\
	&\qquad
	+ w^{\rho\sigma}(Z_\tau) \, dM_\tau^\nu \int_{\tau}^{\tau+d\tau} \, ds
	+ \p_\kappa w^{\rho\sigma}(Z_\tau) \, dM_\tau^\nu \int_{\tau}^{\tau+d\tau} \left(M_s^\kappa - M_\tau^\kappa \right)  ds
	\nonumber\\
	&\qquad
	+ w^{\nu\sigma}(Z_\tau) \, dM_\tau^\rho \int_{\tau}^{\tau+d\tau} \, ds
	+ \p_\kappa w^{\nu\sigma}(Z_\tau) \, dM_\tau^\rho \int_{\tau}^{\tau+d\tau} \left(M_s^\kappa - M_\tau^\kappa \right)  ds
	\nonumber\\
	&\qquad \left.
	+ w^{\nu\rho}(Z_\tau) \, dM_\tau^\sigma \int_{\tau}^{\tau+d\tau} \, ds
	+ \p_\kappa w^{\nu\rho}(Z_\tau) \, dM_\tau^\sigma \int_{\tau}^{\tau+d\tau} \left(M_s^\kappa - M_\tau^\kappa \right)  ds \right]
	+ o(d\tau^2)
	\nonumber\\
	&=
	\E_\tau \Big[
	w^\nu(Z_\tau) \, w^{\rho\sigma}(Z_\tau) \, d\tau^2
	+ w^\rho(Z_\tau) \, w^{\nu\sigma}(Z_\tau) \, d\tau^2
	+ w^\sigma(Z_\tau) \, w^{\nu\rho}(Z_\tau) \, d\tau^2
	\nonumber\\
	&\qquad
	+ \p_\kappa w^{\rho\sigma}(Z_\tau) \int_{\tau}^{\tau+d\tau} \int_{\tau}^s w^{\nu\kappa}(Z_r) \, dr ds
	+ \p_\kappa w^{\nu\sigma}(Z_\tau) \int_{\tau}^{\tau+d\tau} \int_{\tau}^s w^{\rho\kappa}(Z_r) \, dr ds
	\nonumber\\
	&\qquad \left.
	+ \p_\kappa w^{\nu\rho}(Z_\tau) \int_{\tau}^{\tau+d\tau} \int_{\tau}^s w^{\sigma\kappa}(Z_r) \, dr ds \right]
	+ o(d\tau^2)
	\nonumber\\
	&=
	\frac{1}{2} \Big[
	w^{\nu\kappa}(Z_\tau) \, \p_\kappa w^{\rho\sigma}(Z_\tau)
	+ w^{\rho\kappa}(Z_\tau) \, \p_\kappa w^{\nu\sigma}(Z_\tau)
	+ w^{\sigma\kappa}(Z_\tau) \, \p_\kappa w^{\nu\rho}(Z_\tau)
	\Big] \, d\tau^2
	\nonumber\\
	&\quad
	+ \Big[ 
	w^\nu(Z_\tau) \, w^{\rho\sigma}(Z_\tau) 
	+ w^\rho(Z_\tau) \, w^{\nu\sigma}(Z_\tau) 
	+ w^\sigma(Z_\tau) \, w^{\nu\rho}(Z_\tau) \Big] \, d\tau^2
	+ o(d\tau^2)
\end{align}
and
\begin{align}
	\E_\tau \Big[ dZ_\tau^\mu dZ_\tau^\nu dZ_\tau^\rho dZ_\tau^\sigma \Big]
	&=
	\E_\tau \Big[ dM_\tau^\mu dM_\tau^\nu dM_\tau^\rho dM_\tau^\sigma + o(d\tau^2) \Big]
	\nonumber\\
	&=
	\E_\tau \left[ 
	\int_{\tau}^{\tau+d\tau} w^{\mu\nu}(Z_s) \, ds \int_{\tau}^{\tau+d\tau} w^{\rho\sigma}(Z_r) \, dr
	\right. \nonumber\\
	&\qquad
	+ \int_{\tau}^{\tau+d\tau} w^{\mu\rho}(Z_s) \, ds \int_{\tau}^{\tau+d\tau} w^{\nu\sigma}(Z_r) \, dr
	\nonumber\\
	&\qquad \left.
	+ \int_{\tau}^{\tau+d\tau} w^{\mu\sigma}(Z_s) \, ds \int_{\tau}^{\tau+d\tau} w^{\nu\rho}(Z_r) \, dr \right]
	+ o(d\tau^2)
	\nonumber\\
	&=
	\Big[ w^{\mu\nu}(Z_\tau) \, w^{\rho\sigma}(Z_\tau) + w^{\mu\rho}(Z_\tau) \, w^{\nu\sigma}(Z_\tau) + w^{\mu\sigma}(Z_\tau) \, w^{\nu\rho}(Z_\tau) \Big] \, d\tau^2
	\nonumber\\
	&\quad + o(d\tau^2).
\end{align}
If we then use that $w^{\mu\nu}= \alpha \, \lambda \, g^{\mu\nu}$, we find
\begin{align}
	\E_\tau \Big[g_{\mu\nu} \circ dZ_\tau^\mu dZ_\tau^\nu \Big]
	&=
	g_{\mu\nu} w^{\mu\nu} \, d\tau
	\nonumber\\
	&\quad
	+ g_{\mu\nu} \left( 
	w^\mu w^\nu
	+ \frac{1}{2} \, w^\rho \, \p_\rho w^{\mu\nu}
	+ \frac{1}{2} \, w^{\mu\rho} \, \p_\rho w^\nu
	+ \frac{1}{2} \, w^{\nu\rho} \, \p_\rho w^\mu
	+ \frac{1}{4} \, w^{\rho\sigma} \, \p_\rho \p_\sigma w^{\mu\nu}
	\right) d\tau^2
	\nonumber\\
	&\quad
	+ g_{\mu\nu} \Gamma^\mu_{\rho\sigma} \Big( 
	w^\nu w^{\rho\sigma}
	+ w^\rho w^{\nu\sigma}
	+ w^\sigma w^{\nu\rho} \Big) \, d\tau^2
	\nonumber\\
	&\quad
	+ \frac{1}{2} g_{\mu\nu} \Gamma^\mu_{\rho\sigma} \Big(
	w^{\nu\kappa} \p_\kappa w^{\rho\sigma}
	+ w^{\rho\kappa} \p_\kappa w^{\nu\sigma}
	+ w^{\sigma\kappa} \p_\kappa w^{\nu\rho}
	\Big) \, d\tau^2
	\nonumber\\
	&\quad
	+ \frac{1}{4} g_{\mu\nu} \Gamma^\mu_{\rho\sigma} \Gamma^\nu_{\kappa\lambda}
	\Big(
	w^{\rho\sigma} w^{\kappa\lambda} + w^{\rho\kappa} w^{\sigma\lambda} + w^{\rho\lambda} w^{\sigma\kappa} 
	\Big) \, d\tau^2
	\nonumber\\
	&\quad
	+ \frac{1}{3} g_{\mu\nu} \Big( 
	\p_\kappa \Gamma^\mu_{\rho\sigma} + \Gamma^\mu_{\kappa\lambda} \Gamma^\lambda_{\rho\sigma} 
	\Big) 
	\Big(
	w^{\nu\kappa} w^{\rho\sigma} + w^{\nu\rho} w^{\sigma\kappa} + w^{\nu\sigma} w^{\rho\kappa} 
	\Big) \, d\tau^2
	+ o(d\tau^2)\nonumber\\
	&=
	n \, \alpha \, \lambda \, d\tau
	+ g_{\mu\nu} w^\mu w^\nu \, d\tau^2
	+ \alpha\, \lambda \, \Big( 
	\p_\mu w^\mu
	- \Gamma^\mu_{\mu\nu} w^\nu
	\Big) d\tau^2
	\nonumber\\
	&\quad
	+ \frac{\alpha^2\lambda^2}{2} \, g^{\rho\sigma} 
	\Big( 
	g_{\mu\nu} g^{\kappa\lambda} \Gamma^\mu_{\rho\kappa} \Gamma^\nu_{\sigma\lambda}
	+ \Gamma^\mu_{\rho\nu} \Gamma^\nu_{\mu\sigma}
	- \p_\rho \Gamma^\mu_{\mu\sigma}
	\Big) \, d\tau^2
	\nonumber\\
	&\quad
	+ \alpha\, \lambda \, \Big( 
	g_{\mu\nu} g^{\rho\sigma} \Gamma^\mu_{\rho\sigma} w^\nu
	+ 2 \, \Gamma^\mu_{\mu\nu} w^\nu
	\Big)  \, d\tau^2 
	\nonumber\\
	&\quad
	- \alpha^2 \lambda^2 g^{\rho\sigma} \Big(
	g_{\mu\nu} g^{\kappa\lambda} \Gamma^\mu_{\rho\kappa} \Gamma^\nu_{\sigma\lambda}
	+ 2 \, \Gamma^\mu_{\rho\nu} \Gamma^\nu_{\mu\sigma}
	\Big) \, d\tau^2
	\nonumber\\
	&\quad
	+ \frac{\alpha^2 \lambda^2}{4} g_{\mu\nu} g^{\rho\sigma} g^{\kappa\lambda} 
	\Big( 
	\Gamma^\mu_{\rho\sigma} \Gamma^\nu_{\kappa\lambda}
	+ 2 \, \Gamma^\mu_{\rho\kappa} \Gamma^\nu_{\sigma\lambda}
	\Big) \, d\tau^2
	\nonumber\\
	&\quad
	+ \frac{\alpha^2 \lambda^2}{3} g^{\rho\sigma} \Big( 
	\p_\mu \Gamma^\mu_{\rho\sigma} 
	+ 2 \, \p_\rho \Gamma^\mu_{\mu\sigma} 
	+ \Gamma^\mu_{\mu\nu} \Gamma^\nu_{\rho\sigma} 
	+ 2 \, \Gamma^\mu_{\rho\nu} \Gamma^\nu_{\mu\sigma} 
	\Big) \, d\tau^2
	+ o(d\tau^2)\nonumber\\
	&=
	n \, \alpha \, \lambda \, d\tau
	+ g_{\mu\nu} \hat{w}^\mu \hat{w}^\nu \, d\tau^2
	+ \alpha\, \lambda \, \nabla_\mu \hat{w}^\mu d\tau^2
	- \frac{\alpha^2 \lambda^2}{6} \, \mathcal{R} \, d\tau^2
	+ o(d\tau^2).
\end{align}
Plugging this result into eq.~\eqref{eq:QuadVelProc} then yields eq.~\eqref{eq:CondExpQuad}.

\end{document}